\documentclass[useAMS,usenatbib]{mn2e}

\usepackage{graphicx,epsfig,upgreek,subfigure}
\usepackage{longtable,amsmath,amssymb}

\newcommand{\msun}{\mbox{M$_{\odot}$}}
\newcommand{\msol}{\mbox{M$_{\odot}$}}

\newcommand{\rsun}{\mbox{R$_{\odot}$}}
\newcommand{\kms}{\mbox{$\rm{km}\,s^{-1}$}}

\newcommand{\e}[1]{\cdot 10^{#1}}

\def \aj {AJ}
\def \mnras {MNRAS}
\def \apj {ApJ}
\def \apjl {ApJL}
\def \aap {A\&A}
\def \nat {Nature}

\def \apjs {ApJS}
\def \aaps {A\&AS}

\title[SN 2009ip]{SN 2009ip \`a la PESSTO: No evidence for core-collapse yet\thanks{Based on observations collected at the European Organisation for Astronomical Research in the Southern Hemisphere, Chile as part of program 188.D-3003 (PESSTO)}}
\author[Fraser et al.]
{Morgan Fraser$^{1}$\thanks{m.fraser@qub.ac..uk},			
Cosimo Inserra$^{1}$,		     		
Anders Jerkstrand$^{1}$,
Rubina Kotak$^{1}$,\newauthor
Giuliano Pignata,$^{2}$,			
Stefano Benetti$^{3}$,							
Maria-Teresa Botticella$^{4}$,		
Filomena Bufano$^{2}$,\newauthor
Michael Childress$^{5}$,		
Seppo Mattila$^{6}$,				
Andrea Pastorello$^{3}$,					
Stephen J. Smartt$^{1}$,\newauthor		
Massimo Turatto$^{3}$,				
Fang Yuan$^{5}$,				
Joe P. Anderson$^{7}$,			
Daniel D.R. Bayliss,$^{5}$,\newauthor	
Franz Erik Bauer$^{8,9}$
Ting-Wan Chen$^{1}$,
Francisco F{\"o}rster Bur{\'o}n$^{7}$,
Avishay Gal-Yam$^{10}$,\newauthor
Joshua B. Haislip$^{11}$,		
Cristina Knapic$^{12}$,
Laurent Le Guillou$^{13}$,
Sebasti\'an Marchi$^{7}$,\newauthor
Paolo Mazzali$^{3,14,15}$,
Marco Molinaro$^{12}$,
Justin P. Moore$^{11}$,			
Daniel Reichart$^{11}$,\newauthor		
Riccardo Smareglia$^{12}$,
Ken W. Smith$^{1}$,
Assaf Sternberg$^{14}$,
Mark Sullivan$^{16}$,\newauthor	
Katalin Tak\'ats$^{2}$,
Brad E. Tucker$^{5}$,			
Stefano Valenti$^{17,18}$,
Ofer Yaron$^{10}$,\newauthor
David R. Young$^{1}$,	
George Zhou$^{5}$\\				
$^{1}$Astrophysics Research Centre, School of Mathematics and Physics, Queen's University Belfast, Belfast, BT7 1NN, UK \\
$^{2}$Departamento de Ciencias Fisicas, Universidad Andres Bello, Avda. Republica 252, Santiago, Chile\\
$^{3}$INAF-Osservatorio Astronomico di Padova, Vicolo dell'Osservatorio 5,  35122 Padova, Italy\\
$^{4}$INAF-Osservatorio Astronomico di Capodimonte, Salita Moiariello 16, I-80131 Napoli, Italy\\
$^{5}$Research School of Astronomy and Astrophysics, The Australian National University, Weston Creek, ACT 2611, Australia\\
$^{6}$Finnish Centre for Astronomy with ESO (FINCA), University of Turku, V\"ais\"al\"antie 20, FI-21500 Piikki\"o, Finland\\
$^{7}$Departamento de Astronom\'ia, Universidad de Chile, Casilla 36-D, Santiago, Chile\\
$^{8}$Pontificia Universidad Cat\'{o}lica de Chile, Departamento de Astronom\'{\i}a y Astrof\'{\i}sica, Casilla 306, Santiago 22, Chile\\
$^{9}$Space Science Institute, 4750 Walnut Street, Suite 205, Boulder, Colorado 80301\\
$^{10}$Benoziyo Center for Astrophysics, Weizmann Institute of Science, 76100 Rehovot, Israel\\
$^{11}$University of North Carolina at Chapel Hill, Campus Box 3255, Chapel Hill, NC 27599-3255, USA\\
$^{12}$INAF- Osservatorio Astronomico di Trieste, Via G.B. Tiepolo 11, 34143\\
$^{13}$UPMC Univ. Paris 06, UMR 7585, Laboratoire de Physique Nucleaire et des Hautes Energies (LPNHE), 75005 Paris, France\\
$^{14}$Max-Planck Institut f\"ur Astrophysik, Karl-Schwarzschildstr. 1, D-85748 Garching, Germany\\
$^{15}$Astrophysics Research Institute, Liverpool John Moores University ,Liverpool, UK \\
$^{16}$School of Physics and Astronomy, University of Southampton, Southampton, SO17 1BJ\\
$^{17}$Las Cumbres Observatory Global Telescope Network, 6740 Cortona Dr., Suite 102, Goleta, CA 93117, USA\\
$^{18}$Department of Physics, University of California, Santa Barbara, Broida Hall, Mail Code 9530, Santa Barbara, CA 93106-9530, USA\\
}

\date{}

\begin{document}
\maketitle
\clearpage
\begin{abstract}
We present ultraviolet, optical and near-infrared observations of the interacting transient SN 2009ip, covering the period from the start of the outburst in October 2012 until the end of the 2012 observing season. The transient reached a peak magnitude of $M_V$=-17.7 mag, and with a total integrated luminosity of 1.9$\times10^{49}$ erg over the period of August-December 2012. The lightcurve fades rapidly, dropping by 4.5 mag from the $V$ band peak in 100 days. The optical and near infrared spectra are dominated by narrow emission lines with broad electron scattering wings, signaling a dense circumstellar environment, together with multiple components of broad emission and absorption in H and He at velocities between 0.5-1.2$\times10^4$ \kms. We see no evidence for nucleosynthesized material in SN 2009ip, even in late-time pseudo-nebular spectra. We set a limit of $<$0.02 \msol\ on the mass of any possible synthesized $^{56}$Ni from the late time lightcurve. A simple model for the narrow Balmer lines is presented, and used to derive number densities for the circumstellar medium of between $\sim 10^{9}-10^{10}$ cm$^{-3}$. Our near-infrared data does not show any excess at longer wavelengths, and we see no other signs of dust formation. Our last data, taken in December 2012, shows that SN 2009ip has spectroscopically evolved to something quite similar to its appearance in late 2009, albeit with higher velocities. It is possible that neither of the eruptive and high luminosity events of SN2009ip were induced by a core-collapse. We show that the peak and total integrated luminosity can be due to the efficient conversion of kinetic energy from colliding ejecta, and that around 0.05-0.1 \msun\ of material moving at 0.5-1$\times10^4$ \kms could comfortably produce the observed luminosity.  The ejection of multiple shells, lack of evidence for nucleosynthesied elements and broad nebular lines, are all consistent with the pulsational-pair instability scenario. In this case the progenitor star may still exist, and will be observed after the current outburst fades. The long term monitoring of SN 2009ip, due its proximity, has given the most extensive data set yet gathered of a high luminosity interacting transient and its progenitor. It is possible that purported Type IIn SNe are in fact analogues of the 2012b event and that pre-explosion outbursts have gone undetected. 
\end{abstract}


\section{Introduction}
\label{s1}

Stars with a zero-age main sequence (ZAMS) mass of greater than $\sim$8 \msun\ end their lives as core-collapse supernova (CCSNe). The differing masses, metallicities and structures which these massive stars can have at the point of collapse give rise to a tremendous diversity of SN types. Among the most peculiar and extreme are the Type IIn ({\it narrow}) SNe \citep{Sch90}, which are typified by narrow ($\sim$100-500 \kms) emission lines of H, which arise from the collision of the SN ejecta with pre-existing dense circumstellar material (CSM; see \cite{Che94} for an overview of the theory of ejecta-CSM interaction).

Type IIn SNe are intrinsically rare, with relative rates  (i.e. rates as a fraction of the total number of CCSNe) estimated to lie between 4 to 9 per cent (\citealp{Sma09}; \citealp{Smi11}). Their rarity is compounded by their diverse nature; with well studied examples including long-lived events (SNe 2005ip and 2006jd; \citealp{Str12}), events which show Wolf-Rayet features (SN 1998S; \citealp{Liu00}, \citealp{Fas01}), and even a related family of events which show interaction with a He-rich rather than H-rich CSM (\citealp{Pas07,Pas08}). 

In some cases, Type IIn SNe have been linked to Luminous Blue Variables (LBVs). LBVs are characterized by having bright absolute magnitudes ($M_{bol}<-9.5$), irregular eruptions where they brighten by more than one magnitude, and a surface temperature during quiescence of between 12-30 kK \citep{Hum94}. As Type IIn SNe require a dense circumstellar environment prior to explosion, a connection to LBVs -- which shed large amounts of material into their CSM  -- seems natural.

The first connection between LBVs and Type IIn SNe was that  eruptions of the former are often confused with the latter --  a fact which led \cite{Van00} to coin the term ``supernova impostors'' to describe extreme outbursts of massive stars. While SN impostors share the key characteristics of Type IIn SNe, viz. narrow emission lines signaling interaction, the crucial difference is that a SN impostor is a non-terminal event, whereas the progenitor star will not survive a SN explosion. The boundary between Type IIn SNe and impostors has been much debated in the literature (see, for example, SN 1961V \citep{Zwi61}, which was revisited by \citealp{Koc11} and \citealp{Van12}, and the debated SN 1994W-like objects, \citealp{Sol98,Des09,Kan12}).

Direct evidence for an LBV-Type IIn SN link was found by \cite{Gal07}, who identified a source coincident with SN 2005gl in pre-explosion imaging, which was subsequently seen to have disappeared after the SN had faded \citep{Gal09}. The bright absolute magnitude of the progenitor for SN 2005gl could only be matched by an LBV. Similar results from archival imaging have been found for SN 2010jl by \cite{Smi11a}. A physical connection between Type IIn SNe and LBVs was also made by \cite{Tru08}, who observed multiple narrow absorption components for H and He in the spectrum of the Type IIn SN 2005gj, which they attributed to variable mass loss in an LBV progenitor (although we note that \cite{Ald06} have proposed an alternative scenario, where SN 2005gj was a Type Ia SN exploding in a dense CSM). 

In August 2009, SN 2009ip was discovered close to NGC 7259 by the CHASE project \citep{Maz09}. While the object was given an SN designation, it was rapidly realized that it was not a genuine SN, but rather an impostor \citep{Smi10,Fol11}. Fortuitously, the site of SN 2009ip had been observed with the {\it Hubble Space Telescope} prior to the 2009 outburst, enabling \citeauthor{Smi10} and \citeauthor{Fol11} to identity a coincident point source with an absolute magnitude of $F606W$=-9.8; and consistent with a massive star of $\sim$50-60 \msol. While the August 2009 eruption faded, SN 2009ip continued to suffer irregular outbursts over the following three years. \citet[henceforth P13]{Pas13} presented coverage of these eruptions, with a unique set of spectroscopic and photometric data, making SN 2009ip one of the best intensively monitored extragalactic transients to date. Remarkably, P13 found evidence for material accelerated to high velocities ($\sim10^4$ \kms) in September 2011. The detection of fast material over one year prior to the 2012a and 2012b events -- and at a point when SN 2009ip had {\it not} undergone core-collapse -- shows that high velocities cannot be taken as proof of a CCSN.

In July-August 2012, SN 2009ip started a new phase of activity, as first reported by \cite{Dra12}. After an initial outburst which reached a peak absolute magnitude of V=-14 and lasted until around September 25 (termed the ``2012a'' event), SN 2009ip rapidly rebrightened to reach an unprecedented peak magnitude of $R$=-18.5 mag (the ``2012b'' outburst). P13, \cite{Mau13} and \cite{Pri13} all published early photometry and spectroscopy of the eruption, which showed both signs of interaction (narrow lines) and broad absorptions in H and He, with velocities of up to 15000 \kms.

In this paper we present high cadence ultraviolet (UV), optical and near-infrared (NIR) photometry of SN 2009ip, together with optical and NIR spectroscopy, covering the period from the peak of the 2012b event, until the end of the observing season in December 2012. The rest of the paper is organized as follows: in Section \ref{s2} we describe the observations obtained, and the reduction techniques used. In Sections \ref{s3} and \ref{s4} we characterize the photometric and spectroscopic evolution of the SN, and derive a bolometric lightcurve. Section \ref{s5} is devoted to simple modelling of the narrow H emission seen in SN 2009ip, which we use to derive densities for the CSM. Finally, in Section \ref{s6}, we attempt to use our observational data and modelling to understand the evolution of SN 2009ip, and discuss whether the outbursts seen in late 2012 represent the final death of the star as a SN, or rather an example of a non-terminal outburst. In all of the following, we adopt the foreground reddening and distance towards SN 2009ip quoted by P13 and others; $\mu$=31.55 mag, A$_R$=0.05 mag. The distance adopted is also in agreement with that found by \cite{Pot12} from applying a geometric technique (the ``Dense Shell Method'') to SN 2009ip itself.

\section{Observations and data analysis}
\label{s2}

Optical spectroscopic follow-up of SN 2009ip was chiefly obtained with the New Technology Telescope (NTT) + EFOSC2, as part of the Public ESO Spectroscopic Survey of Transient Objects (PESSTO)\footnote{www.pessto.org}. The PESSTO data were supplemented with data from the Telescopio Nazionale Galileo (TNG) + DOLORES, and the Australian National University 2.3m telescope + WiFeS. Details of the instrumental configuration used for each spectrum are listed in Table \ref{tab:opt_spec}. All PESSTO data are also immediately publicly available from the ESO archive; high level data products will be available in June-July 2013. All spectroscopic data presented in this paper will be available for download from the Weizmann Interactive Supernova data REPository (WISEREP; \citealp{Yar12})\footnote{http://www.weizmann.ac.il/astrophysics/wiserep/} upon acceptance. 

All long slit spectra were reduced in the standard fashion; bias and overscan subtracted, flat-fielded using normalized lamp flats, wavelength-calibrated to an arc lamp, and flux-calibrated to a spectophotometric standard star. In the case of the EFOSC2 spectra obtained by PESSTO, these steps were carried out within the PESSTO pipeline\footnote{The PESSTO pipeline has been developed by S. Valenti, and comprises a set of {\sc python} scripts which call {\sc pyraf} tasks to reduce EFOSC2 and SOFI data.}, while for the DOLORES spectra the reduction was performed within {\sc iraf}\footnote{{\sc iraf} is distributed by the National Optical Astronomy Observatory, which is operated by the Association of Universities for Research in Astronomy (AURA) under cooperative agreement with the National Science Foundation.}. The PESSTO data were corrected for telluric absorption by subtracting a model spectrum of the telluric bands, while the DOLORES spectra were corrected using a smooth spectral standard. For spectra where there were a significant number of cosmic ray hits, the {\sc lacosmic} algorithm \citep{Van01} was used to identify and mask cosmic rays in the two-dimensional frames before extraction. The resolutions of the slit spectra were checked from the FWHM of narrow night sky lines, and were found to be $\sim$15 \AA\ and $\sim$10 \AA\ for the EFOSC2 and DOLORES data respectively. These values are in good agreement with the expected resolution from the grating characteristics (reported in Table \ref{tab:opt_spec}, and correspond to a resolution in velocity space (at the wavelength of H$\alpha$) of $\sim$500-700 \kms.

Some optical spectra of SN 2009ip were obtained with the Wide Field Spectrograph \citep[WiFeS;][]{dopita07, dopita10} on the Australian National University (ANU) 2.3 m telescope at Siding Spring Observatory in northern New South Wales, Australia. Observations were obtained with either the B3000/R3000 gratings, which provide wavelength coverage from 3500 to 9600 \AA\ at a resolution of $\sim$2 \AA, or the U/B/R/I7000 gratings which cover 3500 to 9100 \AA\ with $\sim$1 \AA\ resolution. As WiFeS is an integral-field spectrograph, the data were not reduced using the same techniques as for the rest of the optical spectra. Instead, data cubes for WiFeS observations were produced using the {\sc pywifes} software\footnote[1]{{\tt http://www.mso.anu.edu.au/pywifes/}}, and final spectra were obtained with a PSF-weighted extraction routine.

NIR spectroscopy was obtained with the NTT+SOFI and LBT+LUCIFER as detailed in Table \ref{tab:nir_spec}. Solar analogues at a similar airmass were observed either before or after SN 2009ip, to facilitate the removal of the strong telluric absorptions which lie between 1 and 2\micron. For two epochs of NIR spectra (on Nov 13 and Dec 4), a flux standard was observed on the same night and used to calibrate the spectra, for all other data the telluric standard was also used to flux calibrate SN 2009ip. The LUCIFER spectrum was reduced with the ad-hoc pipeline of the LBT Spectroscopic Reduction Center of IASF-Milano, which includes bad pixel, dark and flat-field correction, optimal extraction, sky-subtraction, and wavelength and flux calibration.

Optical photometry was obtained with the NTT + EFOSC2 + $U$\#640,$B$\#639,$V$\#641,$R$\#642,$i'$\#705 (on the same nights as the EFOSC2 optical spectroscopy); with the Panchromatic Robotic Optical Monitoring and Polarimetry Telescope (PROMPT; \citealp{Rei05}) telescopes + $BVRI$ on Cerro Tolodo, and with the Liverpool Telescope + RATCAM + $u'BVr'i'$. In addition, regular photometry of SN 2009ip was obtained with the UltraViolet and Optical Telescope (UVOT) onboard the {\it Swift} satellite (Programme IDs 31486, 32579; P.I. Margutti). The PROMPT and RATCAM images were reduced automatically (trimmed, bias-subtracted and flat-fielded) by their respective pipelines. Significant fringing was evident in the LT RATCAM $i'$ images, and so a fringe frame was constructed from deep exposures of blank fields, scaled to match the science images, and subtracted from them. The EFOSC2 images were reduced using the PESSTO pipeline, again subtracting a smoothed fringe frame in $I$.

For the PROMPT and RATCAM data, point-spread function (PSF) fitting photometry was performed using the {\sc snoopy} package within {\sc iraf}. {\sc snoopy} is a PSF-fitting package using the routines in the {\sc daophot} package, but specifically designed for SN photometry, and including several additional functions such as artificial star tests to estimate photometric errors. The zero point for each image was determined from aperture photometry of the sequence stars listed in P13, while the same colour terms as used by P13 were applied to the instrumental magnitudes. The smaller field of view of EFOSC2, coupled with the short exposure times necessary to avoid saturating the bright target, resulted in only a single reference star bring present in the images obtained in October. As this is insufficient to build a PSF for each frame, we performed aperture photometry on these data. As for the PROMPT and LT data, the zeropoint for each night and filter was determined from local sequence stars, while colour terms for EFOSC2 were evaluated from observations of Landolt standard fields on three photometric nights. All optical photometry, calibrated to the Landolt photometic system is reported in Table \ref{tab:opt}.

The {\it Swift} UVOT frames were reduced using tools within HEAsoft\footnote{available from the NASA High Energy Astrophysics Science Archive Research Center}. Individual images for each epoch were first co-added, before aperture magnitudes were measured following the prescription of \citet{Poo08}. To transform the UVOT {\it ubv} magnitudes from the instrumental system into standard {\it UBV} Johnson magnitudes, shifts of $\Delta U=0.270$, $\Delta B=0.018$ and $\Delta V=-0.042$ mag were applied. These shifts were computed from the magnitudes of the reference stars in the SN field, and are consistent with those reported in P13. These corrections have been applied to all UVOT magnitudes and the resulting magnitude values are reported in Table \ref{tab:opt}. The {\it uvw2, uvm2, uvw1} magnitudes are reported in Table \ref{tab:swift} in the UVOT photometric system.

NIR imaging of SN 2009ip in $JHKs$ were obtained with the NTT + SOFI, the LBT + LUCIFER, and with the Nordic Optical Telescope + NOTCAM. Multiple, dithered, on-source images were taken in each filter, these images were then flat fielded and median combined to create a sky frame. The sky frame was then subtracted from each of the individual images, which were then aligned and coadded. For the LUCIFER imaging, these steps were carried out with the pipeline at the LBT Imaging Data Centre at the Observatory of Rome. Aperture photometry was performed on SN 2009ip, taking care to avoid any flux from the nearby foreground star to the north. Zeropoints for each image were determined from aperture photometry of the same sequence stars as used for the optical photometry, calibrating them to their 2MASS $JHK$ magnitudes. While the $K$ and $Ks$ filters are slightly different, this has a negligible effect on our photometry. NIR magnitudes for SN 2009ip are reported in Table \ref{tab:nir}.

\begin{table}
\caption{Optical spectroscopy of SN 2009ip, together with the nominal resolution for each grating.}
\label{tab:opt_spec}
\begin{tabular}{lccr}
\hline
Date			& Instrument		& Grating			& Res. (\AA)	\\
\hline
2012-09-22	& ANU 2.3m+WiFeS	& B3000+R3000	& 2				\\ 
2012-09-23	& ANU 2.3m+WiFeS	& B3000+R3000	& 2 				\\ 
2012-10-08 	& NTT+EFOSC2	& Gr\#11,16		& 16				\\ 
2012-10-09 	& NTT+EFOSC2	& Gr\#11,16		& 16	  			\\ 
2012-10-14 	& NTT+EFOSC2	& Gr\#11,16		& 16	 			\\ 
2012-10-15 	& NTT+EFOSC2	& Gr\#11,16		& 16	 			\\ 
2012-10-20	& NTT+EFOSC2	& Gr\#11,16		& 16	 			\\ 
2012-10-22	& NTT+EFOSC2	& Gr\#11,16		& 16	 			\\ 
2012-10-22	& TNG+DOLORES	& LRB			& 11				\\ 
2012-10-22	& ANU 2.3m+WiFeS	& B3000+R3000	& 2				\\ 
2012-10-23	& ANU 2.3m+WiFeS	& U,B,R,I7000		& 1				\\ 
2012-10-31	& ANU 2.3m+WiFeS	& B3000+R3000	& 2				\\ 
2012-11-04	& NTT+EFOSC2	& Gr\#11,16		& 16	 			\\ 
2012-11-07	& NTT+EFOSC2	& Gr\#11,16		& 16	 			\\ 
2012-11-12	& NTT+EFOSC2	& Gr\#11,16		& 16	 			\\ 
2012-11-14	& NTT+EFOSC2	& Gr\#11,16		& 16	 			\\ 
2012-11-19	& ANU 2.3m+WiFeS	& B3000+R3000	& 2				\\ 
2012-11-20	& NTT+EFOSC2	& Gr\#11,16		& 16	 			\\ 
2012-12-03	& NTT+EFOSC2	& Gr\#11,16		& 16	 			\\ 
2012-12-09	&TNG+DOLORES	& LRB+LRR		& 11,9			\\ 
2012-12-11	& NTT+EFOSC2	& Gr\#11,16		& 16	 			\\ 
2012-12-20	& NTT+EFOSC2	& Gr\#11,16		& 16				\\ 
\hline
\end{tabular}
\end{table}

\begin{table}
\caption{NIR spectroscopy of SN 2009ip}
\label{tab:nir_spec}
\begin{tabular}{llcc}
\hline
Date			& MJD		& Instrument		& Grating	 \\
\hline
2012-10-16	& 56217.23	& NTT+SOFI		& BG+RG	 \\	
2012-10-22	& 56223.17	& NTT+SOFI		& BG+RG	 \\ 	
2012-10-25	& 56225.11 	& LBT+LUCIFER	& zJ		 \\
2012-11-05	& 56237.11	& NTT+SOFI		& BG+RG	 \\ 	
2012-11-13	& 56245.14 	& NTT+SOFI		& BG+RG	 \\	
2012-12-04	& 56266.08	& NTT+SOFI		& BG+RG	 \\	
\hline
\end{tabular}
\end{table}

\section{Spectroscopic data}
\label{s3}

The full set of spectra collected for SN 2009ip is presented in Tables \ref{tab:opt_spec} and \ref{tab:nir_spec}. In general, the spectra are blue at early times, and are dominated by narrow emission lines from H and He {\sc i}. As the SN evolves, broad ($\sim10^4$\kms) absorptions appear in the Balmer lines, while the [Ca {\sc ii}] NIR triplet emerges in emission, along with Na {\sc i} D. We do not observe a transition to a nebular SN spectrum; indeed the last spectrum in our data set appears to resemble most closely the spectrum of SN 2009ip from September 2009, with the same lines present at both epochs, albeit at higher velocities in 2012. In Sections \ref{s3a} and \ref{s3b}, we describe in detail the spectroscopic evolution of SN 2009ip.

\begin{figure*}
\includegraphics[angle=270,width=1\textwidth]{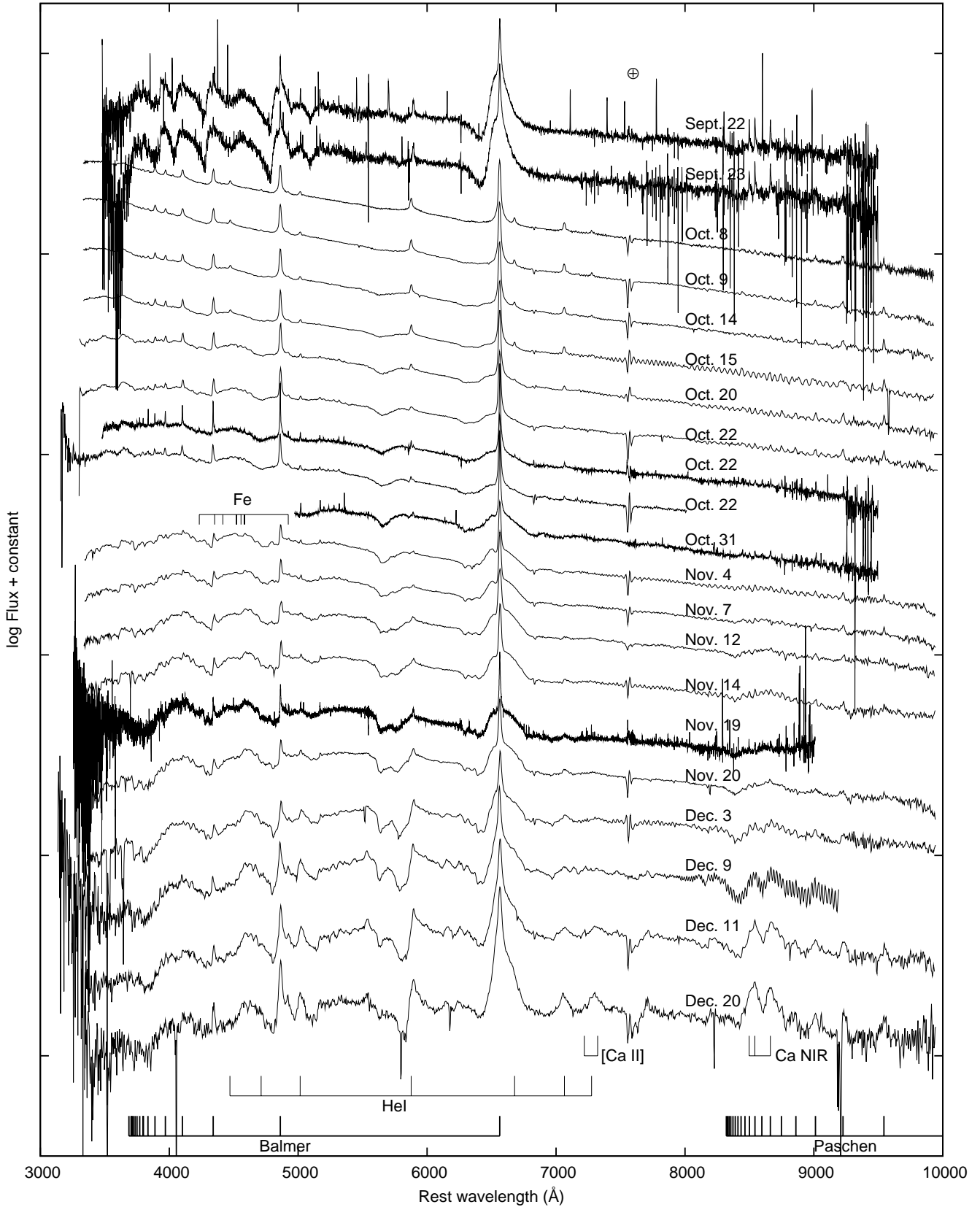}
\caption{The sequence of optical spectra of SN 2009ip, as detailed in Table \ref{tab:opt_spec}. Note that the high resolution spectrum obtained on 23 Oct is not included here, but is shown separately in Fig. \ref{fig:oct23}. The wavelength of residuals from the telluric correction are indicated with \earth.}
\label{fig:allspec}
\end{figure*}

\subsection{Optical spectoscopic evolution}
\label{s3a}

\begin{itemize}

\item {\bf Sept 22 - Sept 23}

The first spectrum in our sequence (from Sept 22) as shown in Fig. \ref{fig:allspec} is contemporaneous with the TNG+DOLORES spectrum presented in P13; coinciding with the end of the 2012a eruption, and immediately prior to the start of the 2012b event. As this spectrum is effectively identical to one obtained on the following night (Sept 23), these two spectra will be discussed together. The most prominent features in the spectra are the Balmer lines. At the rest wavelength of H$\alpha$, SN 2009ip has a narrow (FWHM $\sim$500 \kms) emission component superimposed on broad emission (FWHM $\sim$6000 \kms). A P-Cygni absorption trough is also present with a minimum at -7500 \kms\ with respect to the rest wavelength of H$\alpha$. This absorption is weaker than the broad emission. The higher Balmer lines, notably H$\beta$ and H$\gamma$, have similar multi-component profiles, but with a stronger broad absorption compared to the broad emission. The minimum of the broad absorption in H$\beta$ is also at a lower velocity than that seen in H$\alpha$. A ``shoulder'' of emission is also visible in the blue wing of H$\beta$ and H$\gamma$ (at a velocity of -1800 \kms), which is less visible in H$\alpha$. The He {\sc i} 5876 \AA\ line which appears later is absent at this phase. Metals are present as narrow emission lines, including the blended Na {\sc i} D lines at 5890 \AA\ and 5895 \AA, weak Ca {\sc ii} NIR triplet lines and numerous Fe {\sc ii} lines such as Multiplets 27 and 42.

\begin{figure}
\includegraphics[angle=270,width=0.5\textwidth]{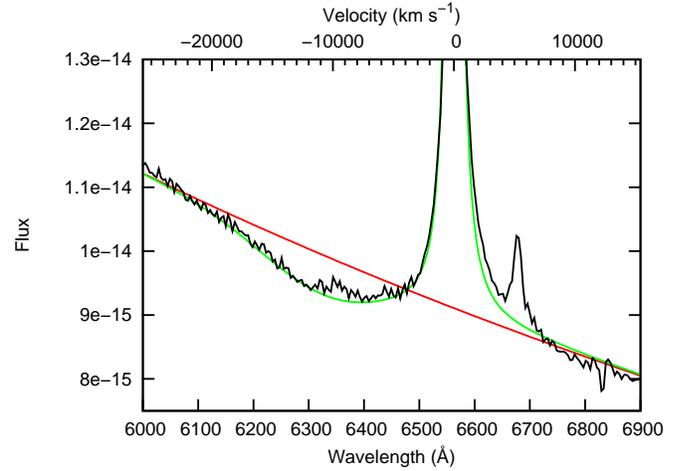}
\caption{The region of H$\alpha$ in our Oct 8 spectrum of SN 2009ip, showing the shallow broad absorption. In red is a blackbody fit to the continuum, while in green is the same blackbody, with a Lorentzian line of FWHM 17 \AA\ and a Gaussian absorption with FWHM 270 \AA\ superimposed. The Lorentzian plus continuum is an excellent fit to the red wing of H$\alpha$ (with the exception of He $\lambda$6678), whereas in the blue wing there is an apparent absorption which is fitted by the Gaussian. As discussed in the text, it appears that the absorption may have two components, although a single Gaussian (as shown here) also gives a reasonable fit. The second x axis shows the velocity with respect to the rest wavelength of H$\alpha$.}
\label{fig:halpha_081012}
\end{figure}

\item {\bf Oct 8 - Oct 15}

The Oct 8 and Oct 9 spectra were obtained when SN 2009ip was at the peak of the 2012b event. The spectra have a blue continuum, and are dominated by narrow (unresolved, $<$700 \kms) Balmer emission. The Paschen series are also visible at wavelengths $>$8500 \AA\, although with much weaker emission. The narrow central component of H$\alpha$, together with the broad base of the line are well fit by a Lorentzian with a FWHM 750\kms. The broad absorption is still present, but much weaker than in the previous epoch. We show in detail the H$\alpha$ line profile in Fig. \ref{fig:halpha_081012}, together with a Lorentzian fit, clearly showing the broad shallow absorption in the blue. Tantalizingly, the profile appears to resemble the multi-component absorption we see at later epochs; and we obtain a best fit to the absorption component with two Gaussians with minima at -6000 and -12000 \kms. However, as the feature is weak we cannot formally exclude a single broad absorption. He {\sc i} $\lambda$5876 is the strongest line after the Balmer series, it has a similar Lorentzian profile to H$\alpha$ (narrow emission on a broad base, together with a hint of broad absorption in the blue). Other weaker He {\sc i} lines are detected at 4471, 4713, 5015, 6678, 7065 \AA\, and tentatively at 7276 \AA. We see no lines from He {\sc ii} (for example at 4686 \AA), nor of the Ca NIR triplet or Na {\sc i} D.

The next spectra in our sequence are from Oct 14 and Oct 15. These spectra are similar to those obtained on Oct 8 and 9, but with a double absorption in the Balmer lines. Fitting two Gaussians to the absorptions in H$\alpha$, we find velocities of -8000 and -12500 \kms. Interestingly, while H$\beta$ and H$\gamma$ appear to have the same double absoption as H$\alpha$, the lower velocity component is at only $\sim$4000 \kms. The higher velocity component is difficult to measure accurately due to the steep continuum, but it too seems to be lower than in H$\alpha$. He {\sc i} 5876 \AA\ has a similar narrow Lorentzian emission component to the preceding epoch, but now also has a clear broad absorption centered on -12500 \kms. This absorption does not appear to have the two-component structure seen in the Balmer lines, but is instead well fitted by a single Gaussian absorption. The broad absorption is not seen in the other He {\sc i} lines, but as these are much weaker than  He {\sc i} 5876 \AA\ this is perhaps not surprising. While an alternative source for the broad absorption could be Na, this would imply a velocity of 13500 \kms, which is uncomfortably higher than H$\alpha$.

\item {\bf Oct 20 - Oct 23}

On Oct 22 we obtained spectra of SN 2009ip from three different telescopes, all of which are similar to that obtained two days earlier on Oct 20. At this phase, the narrow emission on a broad base in H$\alpha$ can no longer be fitted with a single Lorentzian profile. Instead, we find that we need two components to fit the emission; one narrow (FWHM $\sim$800 \kms) and one broad (FWHM $\sim$10000 \kms). The broad emission is not symmetric, as a ``shoulder'' of emission is present in the blue wing of H$\alpha$ which is not present in the red. While the broad absorption can be reasonably fit with a single Gaussian at -10000 \kms, if we take the position of the two absorption components from the previous epoch, then we can see two notches still present at velocities of $\sim$8000 and 12000 \kms. H$\beta$ has a similar broad absorption to H$\alpha$, but with a much less pronounced broad emission, while in H$\delta$ the broad emission is absent. For He {\sc i}, the narrow emission is much weaker than in previous epochs, and there is no strong broad emission, while we still see a strong absorption at -11000 \kms in the 5876 \AA\ line. At this phase, we also see the emergence of a broad emission feature at $\sim$4500  \AA\, with a FWHM of $\sim$10000 \kms. It is difficult to judge the strength of the 4500 \AA\ feature, as it is bracketed by the broad absorption of H$\beta$ and H$\delta$, which serve to lower the apparent level of the continuum. While this feature was present during Oct 14 and 15, it was much weaker, and was plausibly associated with the He 4471 \AA\  line. However, for the 4500 \AA\ line to be caused by He {\sc i} on Oct 22 would require the peak of the emission to be red-shifted by $\sim$2000 \kms. More problematically, it is difficult to see how we could get emission in the 4471 \AA\ line, without corresponding emission in the 5876 \AA\ He line.   While there are no strong, low-ionization state lines arising from any intermediate mass elements around 4500 \AA\, we note that in Type II SNe an emission feature at this wavelength is often seen at early times, arising from multiple blended Fe {\sc i} and Fe {\sc ii} lines such as multiplets 37 and 38. The Fe which gives rise to this in SNe is primordial, so it is entirely possible that we could see a similar feature in a massive non-terminal eruption of an LBV. We also see two emission lines (FWHM $\sim$50 \AA) in the extreme blue end of the spectrum, at 3490 and 3650 \AA. The 3650 \AA\ feature may be caused from the combined flux of the higher order Balmer lines (10--2 transitions and higher), as the series converges to this wavelength, and the blue edge of the line corresponds to the wavelength at which the separation between the lines becomes lower than the instrumental resolution of EFOSC2 (for comparison we can clearly resolve the H$_{J=9-2}$ line in the same spectrum). The identity of the 3490 \AA\ line is uncertain, although as for the $\sim$4500 \AA\ bump, it may be attributable to blended Fe emission.

We obtained a higher resolution spectrum of SN 2009ip on October 23 with WiFeS using the R=7000 gratings (Fig \ref{fig:oct23}). We measured a FWHM of the narrow Lorentzian emission in H$\alpha$ of $266\pm13$ \kms, 243 \kms\ in H$\beta$ and 242 \kms\ in H$\gamma$. As the FWHM is larger than the instrumental resolution of WIFES (45\kms\ at H$\alpha$), we adopt their mean (250$\pm$11 \kms) as the velocity of the CSM. The 3650 \AA\ feature is indeed seen to be comprised of a large number of Balmer lines, as the blue edge of the bump as seen in the spectra from Oct 22 are resolved out into  a blend of the Balmer lines with upper levels between J=13 and J=22. The other striking feature is the appearance of a large number of weak, narrow emission lines from Fe. These features are not seen in the lower resolution data from the preceding night, but were present in spectra of earlier eruptions obtained with XShooter by P13. Unfortunately we do not observe the auroal O line at 5007 \AA\, precluding a direct estimate of the metallicity of the environment.

\begin{figure*}
\includegraphics[angle=270,width=1\textwidth]{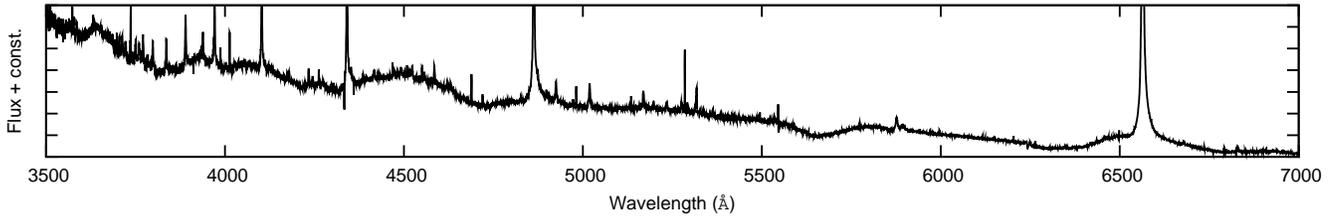}
\caption{The highest resolution spectrum of SN 2009ip in our data set, taken on 23 Oct with the U7000, B7000, R7000 and I7000 gratings on WiFeS, showing the forrest of narrow Fe emission lines.}
\label{fig:oct23}
\end{figure*}

\begin{figure*}
\includegraphics[angle=270,width=1\textwidth]{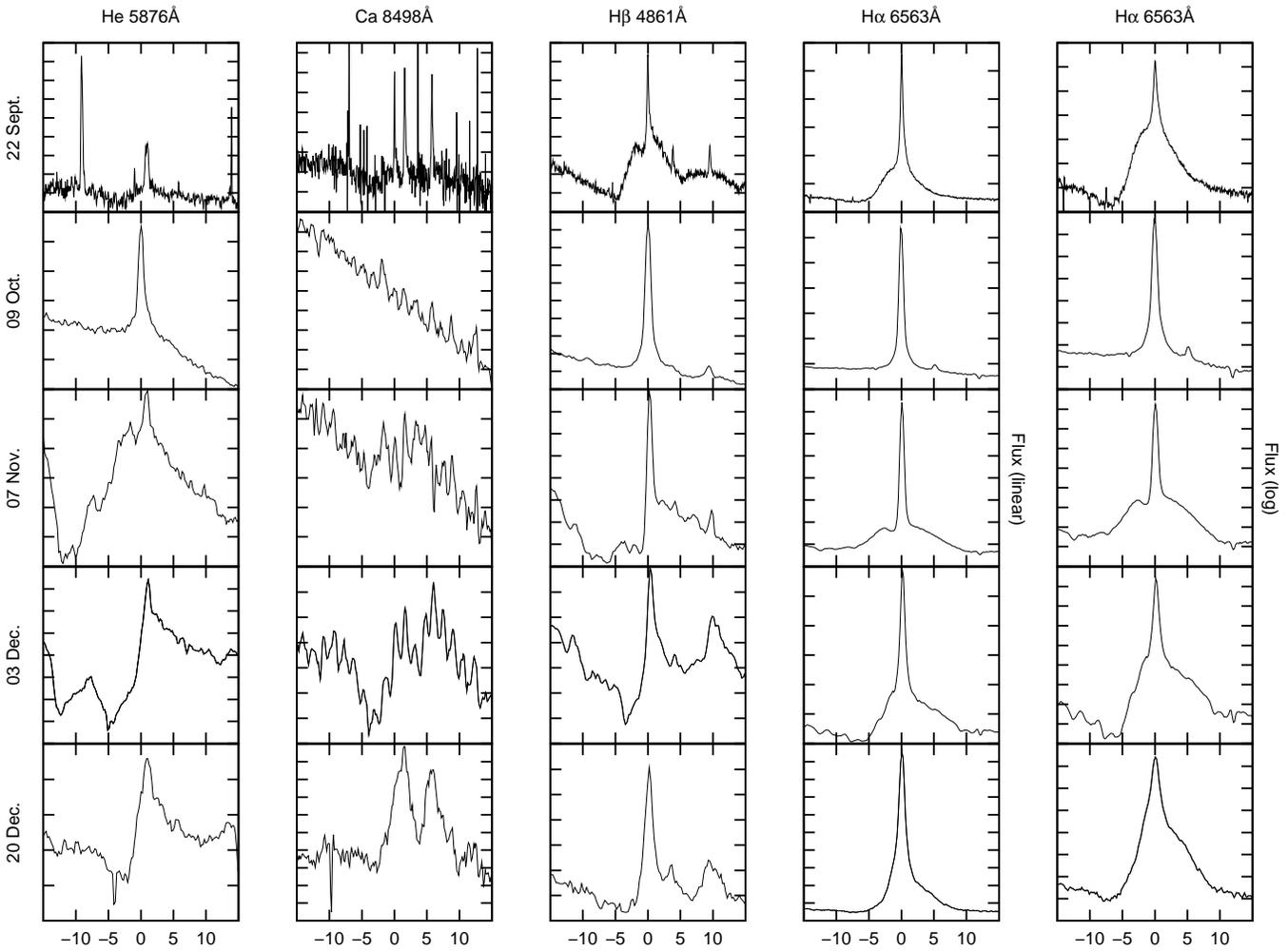}
\caption{The evolution of the line profiles for selected elements in SN 2009ip. All panels are on a common velocity scale, centered on the rest wavelength of the line in question, with units of 10$^3$ \kms. From top to bottom, the rows correspond to: pre-2012a outburst (Sept 22), peak 2012a outburst (Oct 9), first ``bump'' in the lightcurve (Nov 7), approximately two months past peak (Dec 3), and the latest spectrum (Dec 20). All plots are on a linear flux scale, normalized to the peak of the line in question, with the exception of the column on the far right, which shows the H$\alpha$ line on a logarithmic scale to emphasise the structure at the base of the line.}
\label{fig:profiles}
\end{figure*}

\item {\bf Oct 30 - Nov 7}

In the spectra obtained on Oct 30 and Nov 04, H$\alpha$ has quite distinct narrow and broad emission features. The former has a FWHM of $\sim$700 \kms, while the latter has a FWHM $\sim$10000 \kms. The broad absorptions seen at earlier epochs are still present, with multiple components at 12500 and 10000 \kms. We note however that the broad absorptions are weaker than the corresponding broad emission by a factor of a few. The blue shoulder in H$\alpha$ which was beginning to appear in the Oct 22 spectrum is more evident. If caused by an additional emission component, then it consists of material moving at a velocity of -3000 \kms\ with respect to the narrow component, and with a FWHM of $\sim$1500 \kms. Alternatively, the feature could be viewed as an absorption centered on -1500\kms, although this is perhaps less likely, as it would imply that the strong broad emission is intrinsically quite asymmetric. H$\beta$ and H$\gamma$ do not show as strong narrow emission when compared to H$\alpha$. The broad features in H$\beta$ and H$\gamma$ are also quite peculiar. For both lines, the broad emission seen strongly in H$\alpha$ is absent, while the broad absorption is much more pronounced. We see multiple absorptions, with the highest velocity minimum at -12500 \kms in H$\beta$, but with the bulk of the absorption at -8000 \kms. We also see for the first time in the 2012b event clear absorptions from ejecta at quite low velocities, with the first minimum in H$\beta$ at only -2000 \kms. The strong absorption at 5660 \AA\ which we attribute to He {\sc i} with a velocity of -11000\kms\ is still present, without any clear corresponding broad emission. However, the narrow He {\sc i} 5876 \AA\ line has now vanished, and has been replaced with a re-emergent Na {\sc i} D line at 5890,5895 \AA\, suggesting that from this point onwards we cannot be confident that the broad feature is solely caused by He. We also observe the re-appearance of narrow Fe lines in the spectrum of SN 2009ip, such as the 5167 \AA\ line from Multiplet 42.

The spectrum of SN 2009ip from Nov 7 is similar to those discussed in the preceding paragraphs. H$\alpha$ now has a clear, three component absorption, with minima at -12500, -10000 and -6500 \kms. The absorption in H$\beta$ now begins even more sharply at a velocity of only -1000 \kms. Most interestingly however, is the confirmation of a two-component absorption in the line at 5660 \AA. If the line is still dominated by He, then the velocities of the two absorptions are 6000 and 11000\kms, while if Na is dominant, the velocities are slightly higher at 12000 and 7000 \kms. Only the Na {\sc i} D narrow emission is present (albeit weakly), but it is probable that the broad emission is a combination of both He and Na. At this epoch, we also start to see the emergence of a bump around 8500 \AA\ which will develop into the Ca NIR triplet in later spectra. In the blue part of the spectrum, while line identifications are difficult given the complicated velocity structure of the outburst, a broad emission line at $\sim$4100 \AA\ appears to be present, which we tentatively identify as a blend of Fe {\sc ii} lines. 

\item {\bf Nov 12 - Nov 20}

By Nov 12, the ``shoulder'' in the blue wing of H$\alpha$ has begun to fade. We again fit the emission in H$\alpha$ with a narrow and a broad Gaussian, the former is centered on the rest wavelength, while the latter is shifted by 1000 \kms\ to the red, and has a FWHM of $\sim$10000 \kms. The majority of the flux in H$\alpha$ is now emitted by the broad component, while the broad absorptions (which are by now extremely shallow) persist at -12500, -10000 and -6500 \kms. The overall shape of the absorption in H$\alpha$ is flat and shallow, a shape which is mirrored in H$\beta$. We also see the emergence of two emission lines at 5021 and 7075 \AA. While both these features are weak (and so their centers are uncertain), if these are the He {\sc i} 5015 and 7065 \AA\ lines, then they are both systematically redshifted by several hundred \kms. This is in contrast to H$\alpha$, which is within $\sim$1 \AA\ of its rest wavelength. We also see what appears to be a broad bump in the spectrum at $\sim$5250 \AA, to the red of the probable He line. Superimposed on this bump are numerous weak Fe emission lines, indicating that this could be from Fe. In the probable He/Na blend at 5660 \AA, the redder absorption component which emerged on Nov 7 is now comparable in strength to the other absorption. The Ca NIR triplet is clearly emerging in the red part of the spectrum, with high velocities and an apparent P-Cygni profile.

On Nov 20, the complex H$\alpha$ structure persists with absorption up to -12500 \kms, H$\beta$ is largely unchanged from the preceding spectrum. For H$\delta$, the expected absorption is masked by the Fe {\sc ii} line at 4100. The narrow He emission grows in intensity, as does the Ca triplet. Most interestingly, the possible Na/He feature at $\sim$5700   \AA\ now has two clearly separated absorption minima which are well fitted by Gaussians, one centered at -13000 \kms and one at -6500 \kms. The narrow emission is still located at the expected position for the unresolved Na {\sc i} D lines. An apparent absorption feature emerges at 5123 \AA, although at this phase it is very difficult to locate the continuum, and we suggest that this is only an apparent absorption caused by the strengthening of the probable Fe bump between $\sim$5150 -- 5550 \AA, which was first seen in the Nov 12 spectrum.

\item {\bf Dec 3 - Dec 11}

By December 3, the broad base of the H$\alpha$ line has become quite asymmetric, with significantly more flux in the red wing compared to the blue. The narrow emission in H$\alpha$ and H$\beta$ are markedly weaker by this epoch, while H $\delta$ is nearly completely gone.  Interestingly, the highest velocity absorption in H$\alpha$, at $\sim$12,000 \kms\ appear to be weakening. Ca NIR continues to strengthen, with a clear absorption which has a minimum at -3500 \kms\ relative to the bluemost component of the the triplet (8498$\lambda$). We see the emergence of narrow emission features at 6159 and 7313 \AA\ which are probably Fe. 

On December 11, the spectrum is quite similar to that from Dec 3, with the most significant changes being the appearance of [Ca {\sc ii}] at 7219,7323 \AA\, and the re-emergence of strong Paschen lines with a FWHM of $\sim$1500 \kms.

\item {\bf Dec 20}

The last optical spectrum we obtained of SN 2009ip was on Dec 20. We fit the profile of H$\alpha$ with a narrow Lorentzian emission with FWHM 1200 \kms, and a broad Gaussian emission, FWHM 8500 \kms. The centre of the broad component is offset by 800 \kms to the red to reproduce the asymmetric emission. The absorption component is weaker than at previous epochs, and has a blue edge at -11,000 \kms. The [Ca {\sc ii}] emission at 7219,7323 \AA\ continues to grow stronger, becoming comparable in strength to the He {\sc i} 7065 \AA\ line. Meanwhile the P-Cygni profile in the Ca NIR triplet has disappeared, to be replaced with a line purely in emission, which resembles that of a late-time nebular SN line. What is most striking about the final spectrum of SN 2009ip is the resemblance to the spectra of the transient taken three years previously in 2009. A comparison is shown in Fig.\ref{fig:p12_comparison}. All the lines present in September 2009, are also visible in the Dec 2012 spectrum, with the only difference being much higher velocities in the latter. The continuum is somewhat redder in 2012. H, He, Ca, Na and Fe are all present, but we see no evidence for freshly nucleosynthesised materials. There is no sign of [O {\sc i}] at 6300,6364 \AA, and although the emission could be speculated to fill-in some of the absorption in H$\alpha$, we see the same profile in H$\beta$ and H$\alpha$, indicating that this is not the case. Strong Mg {\sc i}] 4571 \AA\ does not appear to be present either, although the presence of Fe emission in this region of the spectrum makes this a less secure result. However, the lack of clear Mg {\sc i}] 4571 \AA\ is in accord with the absence of the 1.5$\upmu$ line in the NIR. We also see the emergence of a line at $\sim$7710 \AA, from December 11 onwards. While the residual from the telluric A-band contaminates this region of the spectrum, the line does appear to be real, however, its wavelength is offset by -60 \AA\ from O {\sc i} 7774. The continuing presence of strong He {\sc i} emission at this epoch is striking, and could be further evidence that interaction is a significant source of energy, as photo-ionization will probably be too weak to support these lines. The strong He lines may also suggest an overabundance of He, although detailed chemical modeling is required to confirm this.

\begin{figure*}
\includegraphics[angle=270,width=1\textwidth]{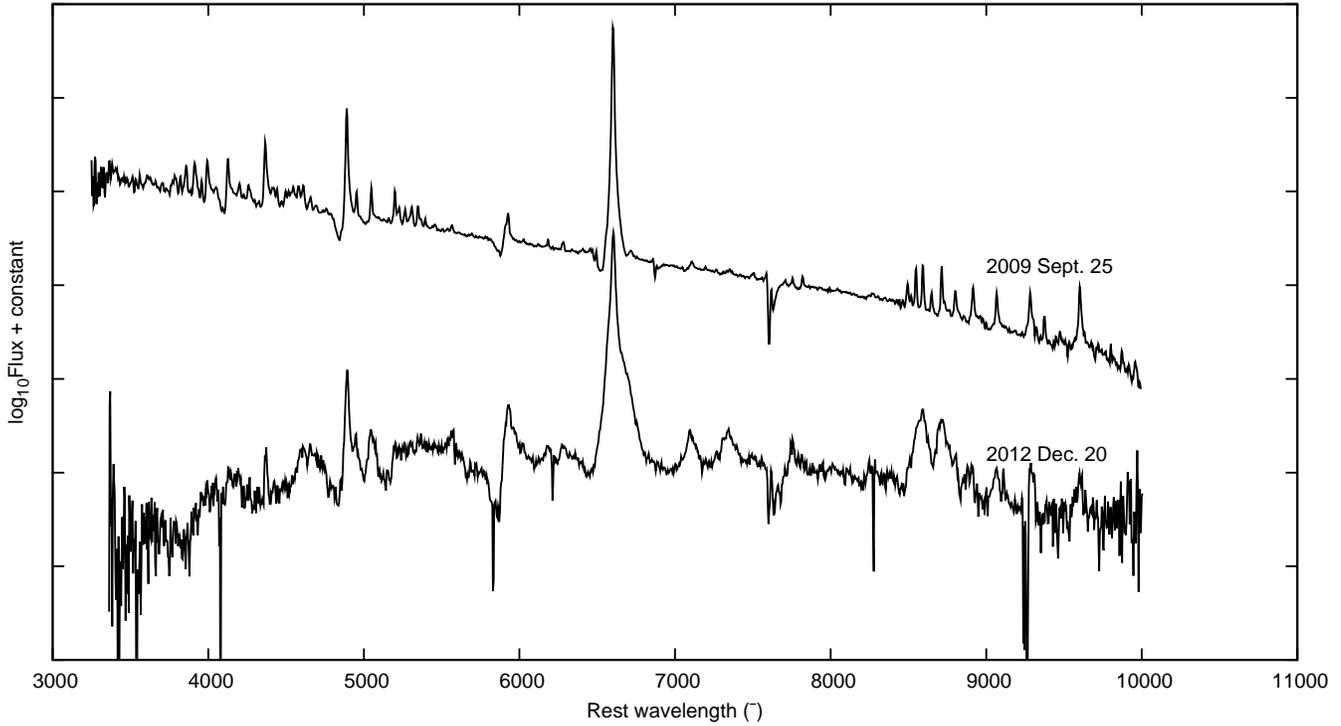}
\caption{Comparison of the September 2009 spectrum of SN 2009ip presented by P13, and the final spectrum we obtained in December 2012. The same lines are present at both epochs, albeit with higher velocities in 2012. In general, the spectra are quite similar, although we note that the Ca and [Fe] emission is more pronounced in 2012.}
\label{fig:p12_comparison}
\end{figure*}

\end{itemize}

\begin{figure}
\includegraphics[angle=270,width=0.5\textwidth]{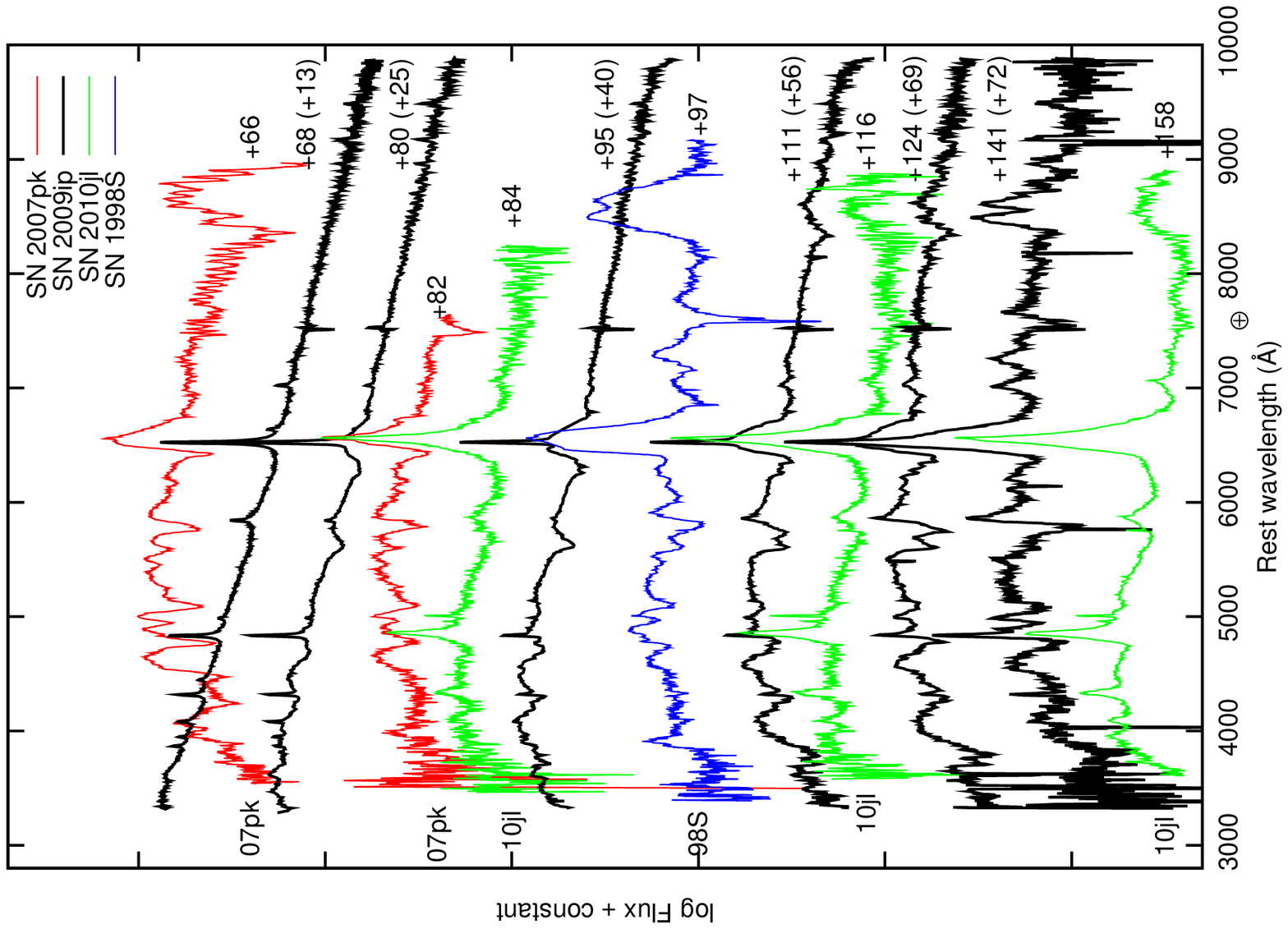}
\caption{Optical spectra of SN 2009ip at representative epochs, compared to a sample of Type IIn SNe. Data are from \protect\cite{Fas01}, \protect\cite{Zha12}, \protect\cite{Ins12}. The phase of each spectrum is indicated in the plot; for the spectra of SN 2009ip the phase from the start of the 2012a event (Aug 1) is given, together with the phase from the start of the 2012b event in parentheses.}
\label{fig:spec_comparison}
\end{figure}

In Fig. \ref{fig:spec_comparison} we compare the spectral evolution of SN 2009ip to that of a sample of interacting SNe, ranging from the weakly interacting Type II SN 2007pk, through to the well studied Type IIn SNe 1998S and 2010jl \citep{Fas01,Zha12,Ins12}. Spectra of SNe 1998S and 2010jl were obtained from the Weizmann Interactive Supernova data REPository (\citealp{Yar12}; WISeREP)\footnote{http://www.weizmann.ac.il/astrophysics/wiserep/}. The best match is the +97 d spectrum of SN 1998S, which is quite similar to the +124 d spectrum of SN 2009ip. However, Ca is much stronger in SN 1998S, while the He lines are weaker compared to SN 2009ip.

\subsection{NIR spectroscopy}
\label{s3b}

\begin{figure*}
\includegraphics[angle=270,width=1\textwidth]{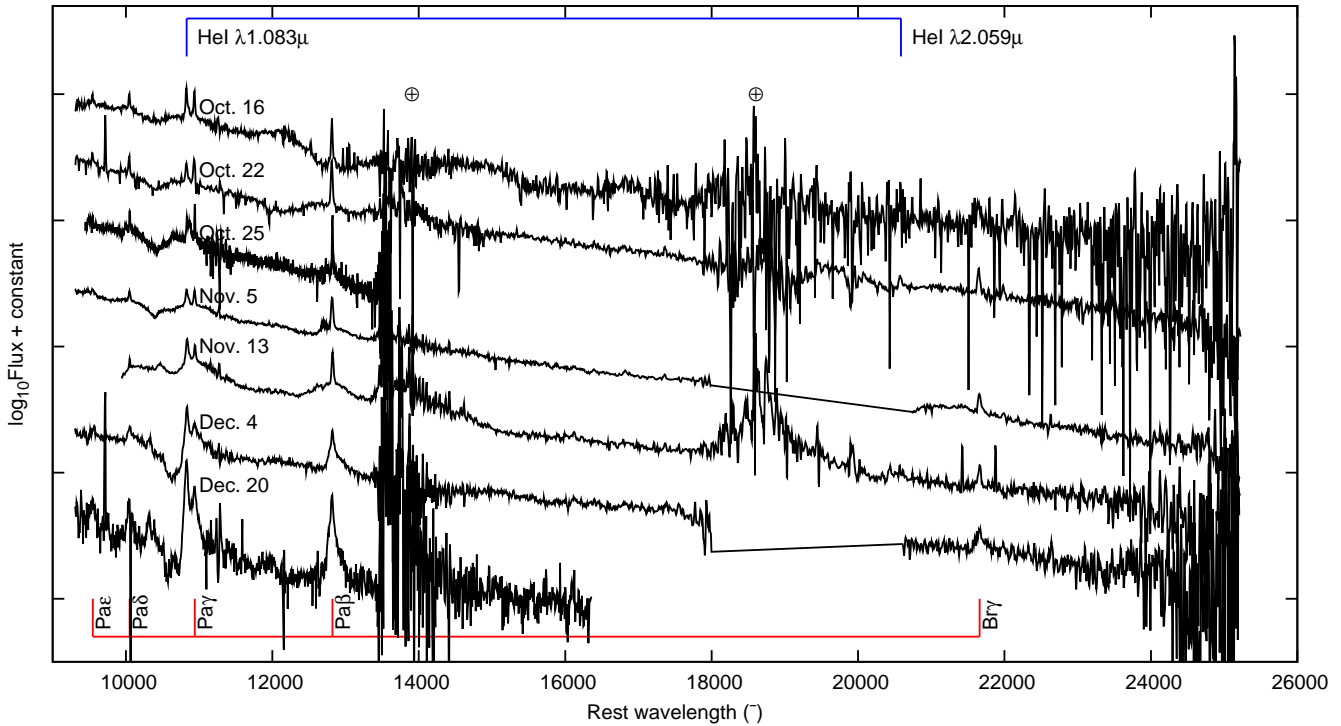}
\caption{The sequence of NIR spectra of SN 2009ip. Regions of strong telluric absorption where the S/N of the spectrum is poor are indicated with a $\earth$ symbol, the rest wavelengths of the H and He lines visible in the spectra are also indicated.}
\label{fig:nir_spec}
\end{figure*}

Aside from the optical spectroscopy, we obtained a sequence of NIR spectra of SN 2009ip as shown in Fig. \ref{fig:nir_spec}. These spectra are dominated by H, with clear detections of the Paschen lines and Br$\gamma$ at all epochs. We also see strong He emission at 1.08$\upmu$m, and more weakly at 2.06$\upmu$m. Besides these elements, there are no other lines present. Most notably, we see no signs of O in the J-band region, while Mg at 1.5$\upmu$m is similarly absent. The spectra do not show any change in their slope over the period between October 16 and December 20, in agreement with the flat $J-H$ and $H-K$ colours seen in Fig. \ref{fig:nir_color}.

\begin{figure}
\includegraphics[angle=270,width=0.5\textwidth]{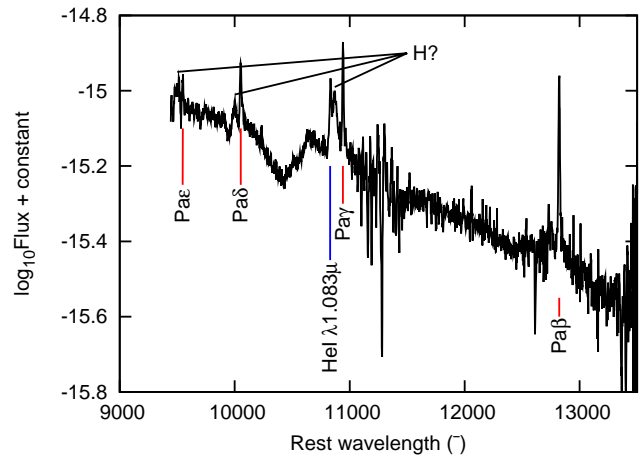}
\caption{Close up of the LUCIFER spectrum described in the text, with the emission components in the blue wing of the Paschen lines indicated.}
\label{fig:lucifer}
\end{figure}

Our highest resolution NIR spectrum is taken with LBT+LUCIFER on October 25, and is shown in detail in Fig. \ref{fig:lucifer}. Pa$\beta$ and $\gamma$ have a FWHM of 330 \kms, slightly higher than the narrow component of the Balmer lines in the optical. Interestingly, we see structure in to the blue wing of the Pa lines,  with emission components at velocities of -1950 \kms\ and -1550 \kms\ relative to Pa$\gamma$ and $\delta$ respectively. This structure, which is not visible in the preceding or following SOFI spectra, is contemporaneous with the emergence of the blue shoulder in H$\alpha$ in the optical. We note however, that the blue shoulder at in the optical spectra was present at velocities of $\sim$3000 \kms, higher than what is seen in the NIR. We also find evidence for a weak P-Cygni absorption in Pa$\delta$ and possibly Pa$\beta$, with minima of -3000 \kms\ for the former, but only -900 \kms\ for the latter. The minimum in Pa$\beta$ is much lower than that observed in other lines, and is comparable to the typical velocities seen in LBV outbursts. We must caution however, that the S/N is not as good here as at shorter wavelengths, and so the identification of this component is more tenuous. 

We see a broad absorption at 1.043$\upmu$, which has a velocity minimum of 14500 \kms\ with respect to Pa$\gamma$. We do not see a corresponding minimum in Pa$\beta$ or $\delta$, and so consider it more likely that this is associated with the strong He {\sc i} line at 1.083 $\upmu$, in which case the absorption minimum lies at 11500 \kms. This is similar to what we measure in the Na/He blend in our optical spectra from October 30.

\subsection{The line profiles of SN 2009ip}
\label{s3c}

The H and He lines in SN 2009ip are characterized by narrow emission on a broad base, and multiple broad absorptions at velocities on order of 10$^4$ \kms. Below we discuss each component of the Balmer lines in turn. 

\begin{itemize}

\item {\bf Narrow H emission}

Similar to what is observed in an H {\sc ii} region, the narrow lines arise from the CSM. While the H in an H {\sc ii} region is ionized from massive OB stars, in the case of Type IIn SNe the CSM is ionized by either the flux of the SN and/or the emission of the cool dense shell. The velocities in the narrow lines ($\sim$500 \kms) are similar to that seen in many Type IIn SNe and LBVs \citep{Kie12}, although lower than those seen in some extreme examples of the class of SN impostors \citep{Pas10}. The narrow lines are present at all epochs in SN 2009ip, although they are decreasing in strength with respect to the broad component over the period of the 2012b event. We see no narrow P-Cygni absorption in H$\alpha$ from a wind, as shown in Fig. \ref{fig:halpha}. This is in contrast to the sample of interacting SNe studied by \cite{Kie12}. 

We measured the central wavelength of the narrow component in H$\alpha$ in all our spectra by fitting a Gaussian to the line, taking care to exclude the broad base from the fit. A plot of central wavelength as a function of time is shown in Fig. \ref{fig:halpha_cent}, which reveals a small but significant trend towards redder wavelengths with time.

\begin{figure}
\includegraphics[angle=270,width=0.5\textwidth]{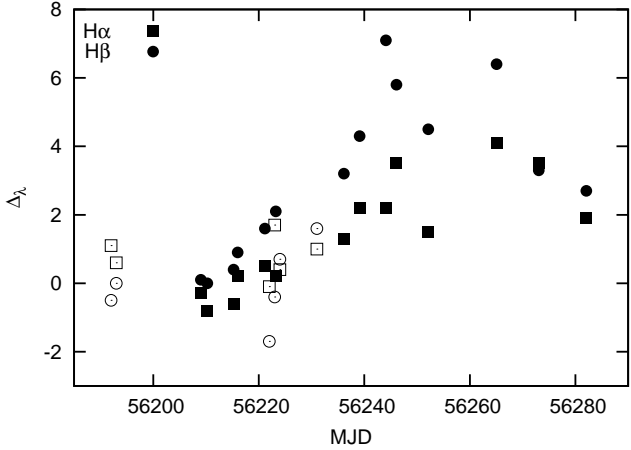}
\caption{Wavelength shift of the narrow components of H$\alpha$ and H$\beta$, relative to the rest frame of NGC 7259. Filled points indicate measurements obtained from the EFOSC2 spectra, which were pipeline reduced in a homogeneous fashion and had their wavelength calibration checked against night sky lines. Open symbols indicate measurements made on spectra from TNG+DOLORES and the ANU 2.3m+WiFeS.}
\label{fig:halpha_cent}
\end{figure}

The slight shift of the narrow component of H$\alpha$ to redder wavelengths is unusual, although we note that a similar effect was seen by \cite{Kan12} in SN 2009kn. A shift to the {\it blue} could be easily explained by the formation of dust in the ejecta. While one may invoke dust destruction to explain the opposite effect, the shift to the red is quite gradual, and does not appear to coincide with the peak of the UV flux which might be expected to destroy the dust. Furthermore if there was also significant dust being destroyed, a NIR infrared excess which disappears over time would also be expected, although we do not observe this in our photometry. Dust formation in a clump of material moving towards the observer could be an alternative explanation, although we do not see  a strong NIR excess in our photometry which may be associated with dust formation.
 
\begin{figure}
\includegraphics[angle=270,width=0.5\textwidth]{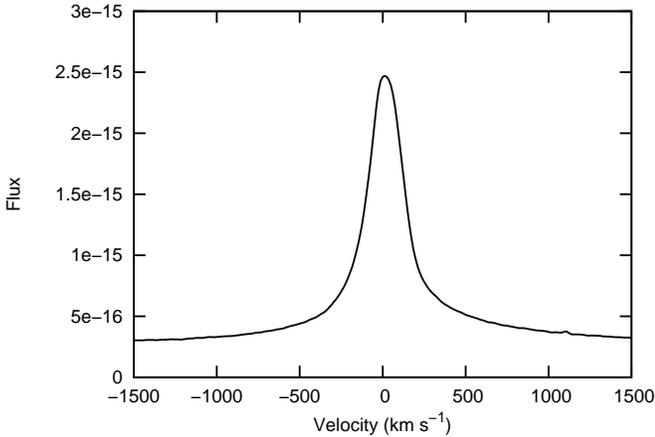}
\caption{H$\alpha$ from our highest resolution spectrum, obtained on Oct 23. We see no evidence for a narrow ($\lesssim$ 250 \kms) P-Cygni absorption at high resolution.}
\label{fig:halpha}
\end{figure}

\item {\bf Broad emission} The broad emission is approximately centered on the rest wavelength of H$\alpha$. While broad emission is usually associated with emission from ejecta at a large velocity, \cite{Chu01} first suggested for the Type IIn SN 1998S that an alternative explanation could be scattering of photons; a similar result was found by \cite{Des09} for SN 1994W. Repeated Thomson scattering of photons by electrons in the dense CSM can give rise to broad lines which have a Lorentzian profile, which can be seen most clearly in the Oct. 8 spectrum of SN 2009ip. An alternative mechanism to create broad lines is boxy emission from interaction, as noted by \cite{Che94}.

The broad emission becomes notably asymmetric after November 4, approximately coinciding with the first bump in the lightcurve. This raises the possibility that the bump is due to the ejecta encountering a denser region of CSM expelled by the progenitor star. If the density is not uniformly enhanced in all directions, then we may expect asymmetries to arise in the profile of H$\alpha$.

\item {\bf Broad absorption}. We see broad absorption extending out to a velocity of -16,000 \kms, with a flat bottomed profile. While a shallow, flat-bottomed absorption trough is unusual in SN and LBVs, it can be explained by the presence of a thin shell of absorbing material \citep{Jef90}. Interestingly, \citeauthor{Jef90} find that in such an arrangement the maximum absorption in the line is 20 per cent of the flux in the continuum, which is in good agreement with what we see in the H$\alpha$ trough in the EFOSC2 spectrum from November 14.

The presence of the broad H$\alpha$ absorption in the spectrum at peak (Fig. \ref{fig:halpha_081012}) indicates that the optical depth of the cool dense shell cannot be much greater than a few. Alternatively, we require a clumpy CSM-ejecta interaction, where the high velocity ejecta is visible through windows in the cool dense shell, and which could be caused by instabilities in the ejecta \citep{Che92}.

A final interesting point of note in SN 2009ip is that the broad absorption appears stronger in H$\beta$ than H$\alpha$. This is in contrast to what is expected, given that the optical depth in H$\alpha$ is always higher than in H$\beta$. A possible explanation for this could be broad H$\alpha$ emission ``filling in'' some of the absorption in H$\alpha$. At high densities, H$\beta$ is quenched as the n=4 level is depopulated to n=2 via Pa$\alpha$. In this case, one would also expect to see strong Pa$\alpha$ emission, but as this line is in the telluric band at 1.87$\upmu$, it is difficult to test this hypothesis.

\end{itemize}

\section{Photometry}
\label{s4}

\subsection{Lightcurve evolution}
\label{s4a}

\begin{figure*}
\includegraphics[angle=270,width=1\textwidth]{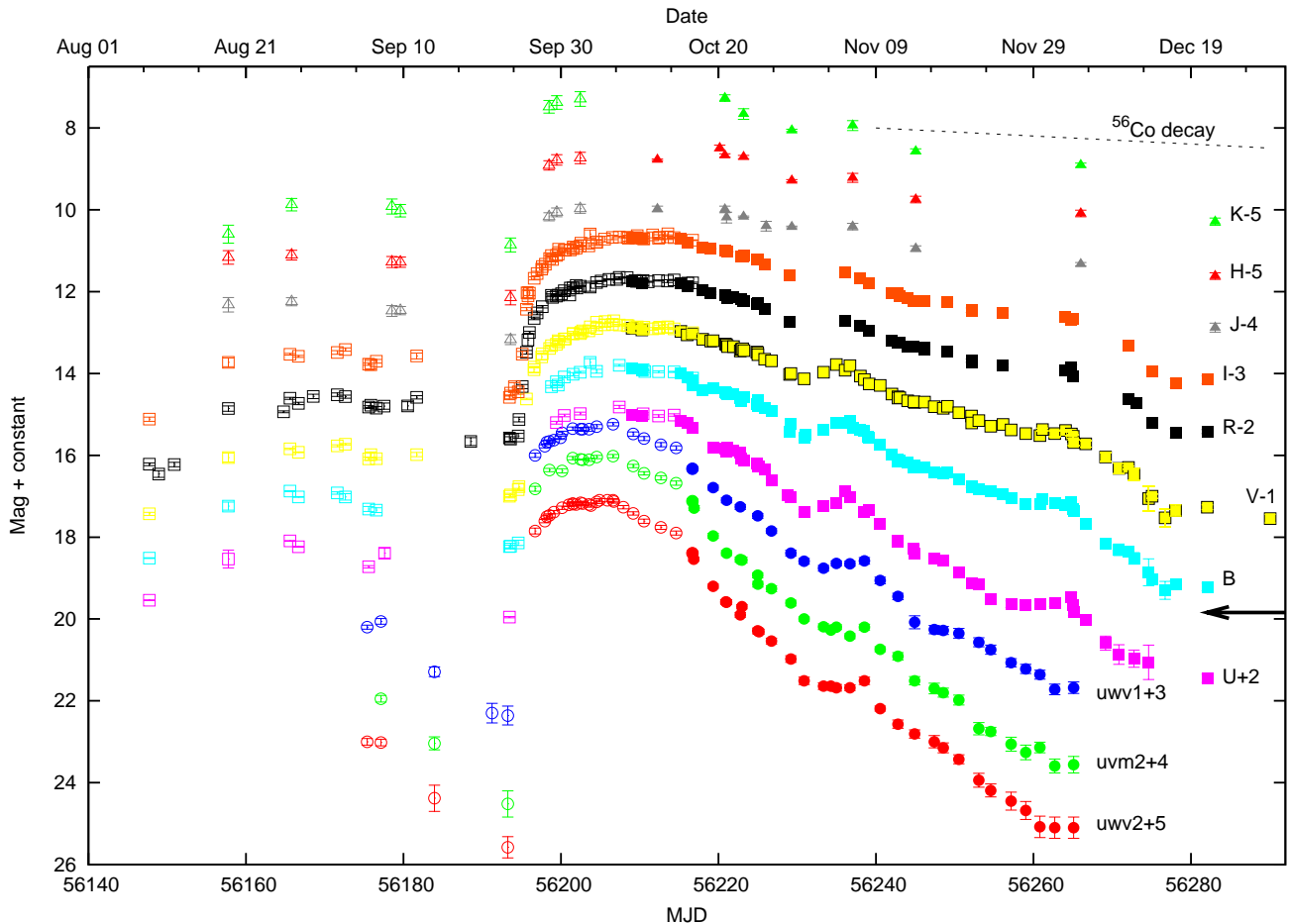}
\caption{Optical (square points), UV (round points) and NIR (triangle points) photometry of SN 2009ip for the 2012a and 2012b events. Open points are taken from the data published by P13, while filled points are new data presented in this paper. We also include the $JHK$ magnitudes reported for Oct 29 by \protect\cite{Im12}. Also indicated are the quiescent $F606W$ ($R$) magnitude of the progenitor from \protect\cite{Fol11} (solid arrow, offset by -2 mag to match the $R$-band magnitudes of SN 2009ip), and the expected 0.98 mag per 100 days decline rate in $V$ given by the radioactive decay of $^{56}$Co (dashed line).}
\label{fig:lc}
\end{figure*}

The photometry of the 2012a event, together with the initial rise of the 2012b event have been presented by P13. There is some overlap between these data, and our first photometry obtained from NTT+EFOSC2. We find excellent agreement with the magnitudes reported by P13 in the region of overlap (Fig. \ref{fig:lc}); both sets of photometry agree to within their uncertainties, and show no systematic offsets. We have also compared our lightcurve to the $BRI$ data presented by \cite{Mau13}, and find in general fair agreement between the two data sets within the uncertainties, and again see no systematic offset.  The region of the bump in the lightcurve (around Nov 7) is the only discrepancy, where \citeauthor{Mau13} find a similar increase in $I$, a larger increase in $B$, and no increase in $R$, although the reason for this disagreement is unclear.

The lightcurve in all filters exhibits a decline from the bolometric maximum of the 2012b event (around October 3, MJD 56203), until the end of the observing season on December 29 (MJD 56290) when SN 2009ip disappeared behind the Sun. The decline is the strongest in the UV bands, with a drop of 8 mag from peak in $uvw2$ over 50 days, compared to only a 2 mag drop over the same period in $I$. During our last observation of SN 2009ip, taken on December 29 in $V$, the transient had faded to $M_V$=-13.1, 5 magnitudes below peak. Even at this phase the SN is still ~4.5 mag brighter in $R$ than the quiescent magnitude reported by \cite{Fol11}, $F606W$=21.84$\pm$0.17.

\begin{figure}
\includegraphics[angle=270,width=0.5\textwidth]{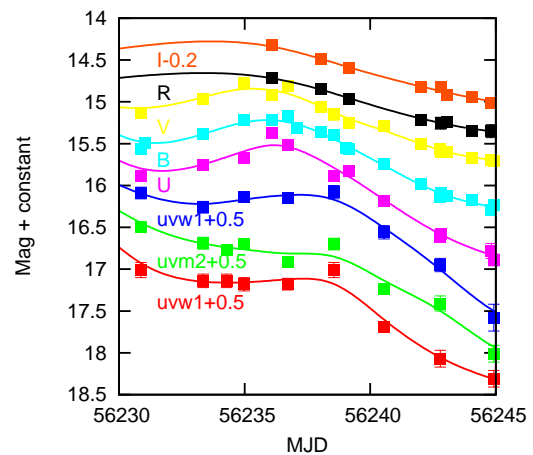}
\caption{The region of the first bump in the lightcurve of SN 2009ip (around Nov 1), along with spline fits to the data.}
\label{fig:peak}
\end{figure}

Around November 1, we see a clear bump in the lightcurve of SN 2009ip, where the transient brightens by 0.2-0.5 mag in all filters. Curiously, the bump appears to develop first in the redder filters. We show a zoomed in section of the lightcurve of SN 2009ip in Fig. \ref{fig:peak}, where we have fitted natural splines to the photometric points between MJD 56230 and 56245 in each filter (taking the points immediately prior and subsequent to this range as  the boundary values, and weighting all points equally). From the spline fits it appears that the bump becomes visible first in $R$ and $I$, and lags by a few days in the {\it Swift} UV filters. The bump also appears to be more pronounced in the optical; while we have probably missed the peak in $RI$, $UBV$ show a clear rise, as opposed to $uvw1,uvm2,uvw1$ which merely flatten for a few days.

A second bump in the lightcurve can be seen at the start of December (MJD $\sim$56265). This bump is less pronounced than the one seen around November 1, with the optical bands merely flattening for a period of around a  week, rather than increasing in brightness. Unfortunately the second bump coincides with the point at which SN 2009ip is too faint to observe in the UV with {\it Swift},  while the $R$ and $I$ bands are sparsely sampled to clearly identify the peak of the bump. The December bump lasts for a similar duration to  the one in November - perhaps a week or less - and is followed by a continuing decline in the lightcurve afterwards. From around December 17 (MJD 56278) the lightcurve appears to flatten until the last epoch on December 29.

\begin{figure}
\includegraphics[angle=270,width=0.5\textwidth]{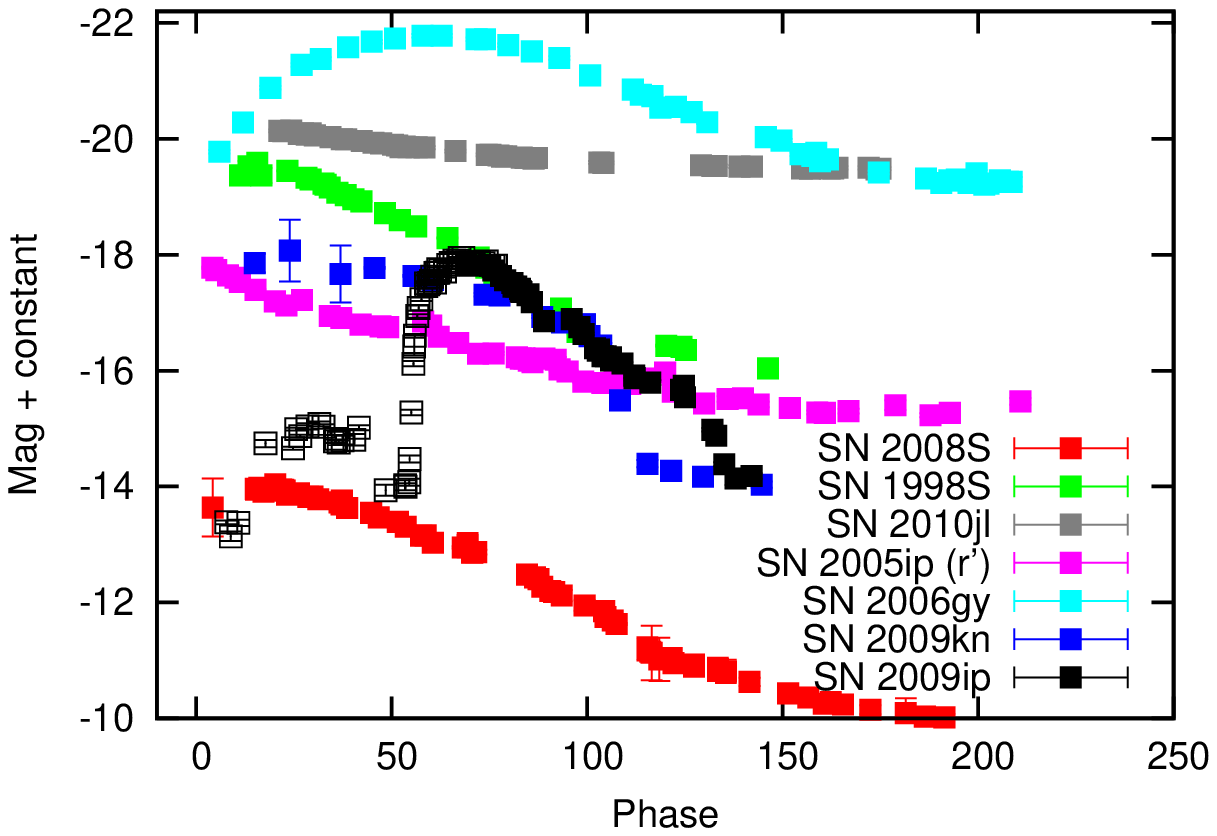}
\caption{Comparison of the absolute R-band lightcurve of SN 2009ip with a selection of interacting transients (SN 2008S, t$_0$=54485.5, A$_R$=1.6, $\mu$=28.7, \protect\citealp{Bot09}; SN 1998S, t$_0$=50874.7, A$_R$=0.56, $\mu$=31.15, \protect\citealp{Fas00}; SN 2010jl, t$_{max}$=54887.5, A$_R$=0.06, $\mu$=33.44, \protect\citealp{Zha12}; SN 2005ip, t$_{disc}$=53679.2, A$_R$=0.011, $\mu$=32.71, \protect\citealp{Str12}; SN 2006gy, t$_{0}$=53976.0, A$_R$=1.68, $\mu$=34.32, \protect\citealp{Smi07}; SN 2009kn, t$_0$=55115.5, A$_R$=0.25, $\mu$=34.23, \protect\citealp{Kan12}). Lightcurves are plotted with respect to the explosion epoch (t$_0$), or where a good estimate of this is unavailable, to the discovery date (t$_{disc}$) or epoch of maximum light (t$_{max}$). Distances and extinctions are taken from the literature.}
\label{fig:r}
\end{figure}

In Fig. \ref{fig:r} we compare the $R$-band lightcurve of SN 2009ip over August -- December 2012 to a sample of interacting transients taken from the literature. The peak magnitude of SN 2009ip, $\sim-18$, is comparable to that of SNe 2005ip and 2009kn, but much fainter than that of Type IIn SNe such as SN 2010jl and 1998S. SN 2009ip also has the most rapid decline from peak of any objects in the sample.

\begin{figure}
\includegraphics[angle=270,width=0.5\textwidth]{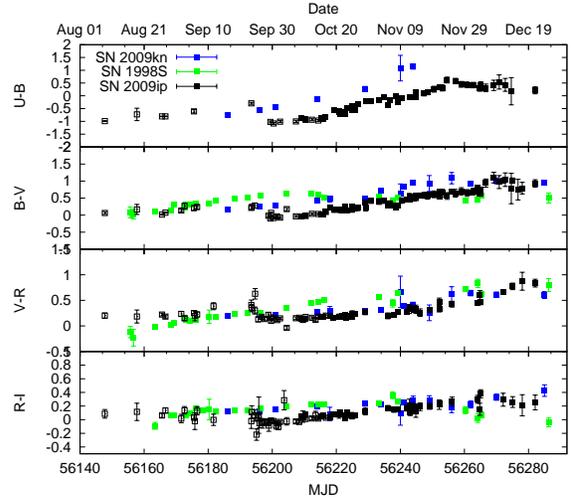}
\caption{Optical colour curves for SN 2009ip during the 2012a and 2012b events. Open symbols are from P13, while closed symbols are from this paper. SNe 2009kn and 1998S \citep{Kan12, Fas00} are also shown, these objects have been shifted so that their explosion epoch is at MJD=56140.}
\label{fig:opt_color}
\end{figure}

\begin{figure}
\includegraphics[angle=270,width=0.5\textwidth]{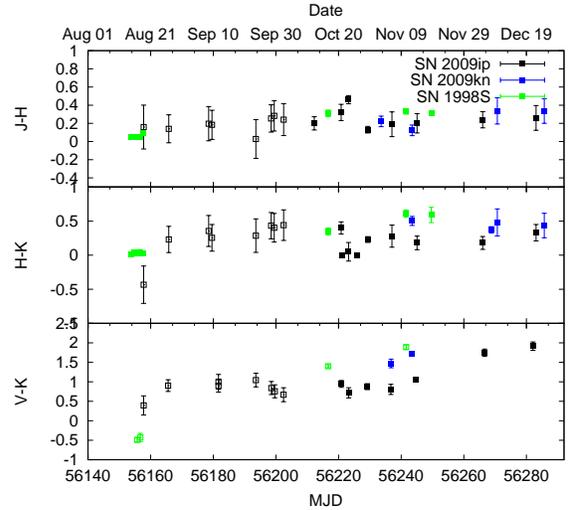}
\caption{NIR colour curves for SN 2009ip during the 2012a and 2012b events. Open symbols are from P13, while closed symbols are from this paper. We also include the $JHK$ magnitudes reported for Oct 29 by \protect\cite{Im12}. SNe 2009kn and 1998S are also shown for comparison \citep{Kan12, Fas00}, shifted so that their explosion epoch would be at MJD=56140. All colour curves have been corrected for reddening.}
\label{fig:nir_color}
\end{figure}

The colour evolution of SN 2009ip is presented in Fig. \ref{fig:opt_color}. For the 2012a event, and up to the peak of the 2012b event, the colour in all optical bands is flat. As the SN fades from the peak of the 2012b event the colour becomes redder in all bands as the photosphere cools. The colour evolution from photometry is in accord which that expected from the spectra, which show a similar slope until peak. The $B-V$ colour shows a bump around the same time as the Nov 7 undulation in the lightcurve, which is to be expected given the time difference between the bump in different filters, as discussed in Section \ref{s4b}. A comparison to other interacting transients (SNe 1998S and 2009kn) shows a good match with SN 2009ip, with the only notable difference being a bluer {\it U-B} color for the latter compared to SN 2009kn. The NIR colours of SN 2009ip are shown in Fig. \ref{fig:nir_color}. As in the optical, we see a similar color evolution to SNe 2009kn and 1998S. The evolution of SN 2009ip is consistent with a cooling blackbody, and shows no evidence for a NIR excess which would signal dust formation.

\begin{figure}
\includegraphics[angle=270,width=0.5\textwidth]{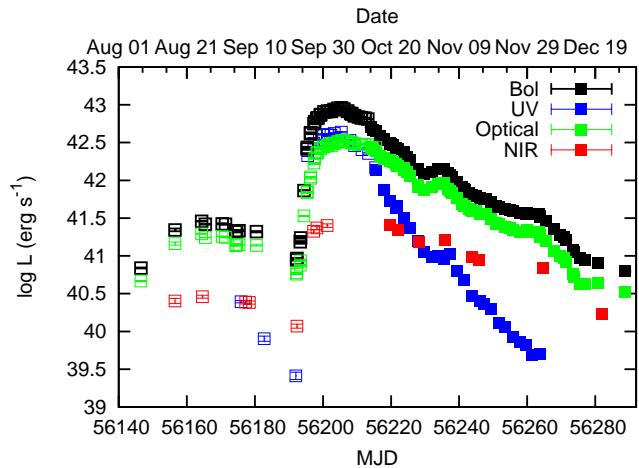}
\caption{The $uvoir$ bolometric lightcurve of SN 2009ip (black), together with the contributions of the optical, UV and NIR photometry}
\label{fig:bol}
\end{figure}

We constructed a bolometric lightcurve for SN 2009ip from our UV, optical and NIR photometry. The measured magnitudes were converted into fluxes at the effective wavelengths of the filter used and corrected for extinction, before the resulting spectral energy distribution was integrated over the wavelengths covered with a linear interpolation, assuming zero flux beyond the limits of the integration. Fluxes were computed at epochs where there was an observation in $V$; where observations in other bands were not taken contemporaneously, their magnitudes were estimated by interpolating the light curves using low-order polynomials, or were extrapolated assuming constant colours. The bolometric flux at each epoch was converted to a luminosity given the adopted distance to SN 2009ip. The resulting bolometric lightcurve is shown in Fig. \ref{fig:bol}. The bolometric lightcurve of the 2012a event is dominated by optical light. For the rising phase of the 2012b event however, the contribution to the lightcurve from the UV is larger than from optical wavelengths. The NIR contributes very little flux to the lightcurve of SN 2009ip for almost the entire duration of the 2012a and 2012b outbursts, amounting to only 20 per cent of the total at the time of our last epoch of NIR photometry.

\begin{figure}
\includegraphics[angle=270,width=0.5\textwidth]{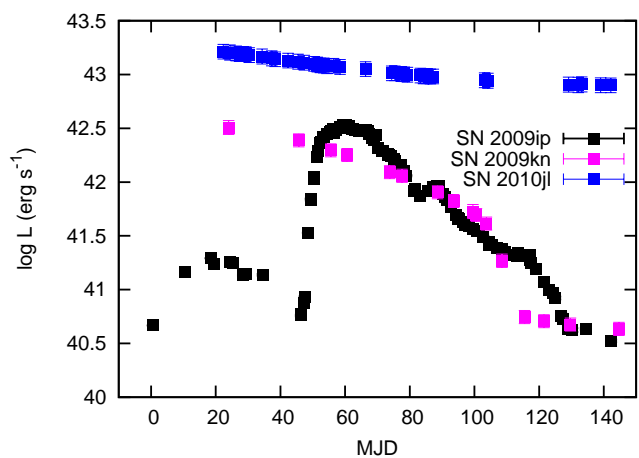}
\caption{The $uvoir$ bolometric lightcurve of SN 2009ip (black), together with two other well observed Type IIn SNe; SN 2009kn \protect\citep{Kan12} and SN 2010jl \protect\citep{Zha12}. The lightcurves of SNe 2009kn and 2010jl are relative to their explosion date, while that of SN 2009ip is relative to the start of the 2012a event.}
\label{fig:bol_comparison}
\end{figure}

\subsection{The undulations in the lightcurve of SN 2009ip}
\label{s4b}

The presence of bumps in the lightcurve of SN 2009ip is unusual (but not unprecedented, as they were also noted in the Type IIn SN 2005la; \citealp{Pas08}). To reproduce these short bumps with the interaction of a few solar masses of ejecta (as might be ejected in a Type IIn SN) with the CSM may be difficult, as a massive SN ejecta should be more homogeneous and not give rise to small scale variations in the lightcurve. It is also possible that the bumps are periodic, although monitoring over a longer timescale is needed to confirm this. In this case, the bumps could possibly be caused by spiral density structures resulting from a passage of a binary companion. 

Another surprising aspect of the bumps is that they appear to peak first in the redder bands. Dust is unlikely to be the cause of the lag in the UV, as while it may cause the bump to be less pronounced at bluer wavelengths, there is no obvious reason why the peaks of the lightcurve in different filters would not coincide.

An alternative explanation for the shift in peak with wavelength is successive scattering by metal lines in the UV. The peak in $U$ and the {\it Swift} UV filters is approximate 1-2 days after the peak in BVRI. To explain this delay with an additional travel time for UV photons implies that they cover a distance of 170-340 AU {\it more} than the optical and NIR photons. For a characteristic radius of ~$\sim$40 AU to the cool dense shell, this requires each photon undergoes multiple scattering events. It is also important that the photons (or at least a high percentage of them) scatter from the lines (which will give rise to re-emission at the same wavelength), rather than causing the atom to fluoresce (which would shift flux towards redder wavelengths) or undergo collisional de-excitation.

There is no marked change in the spectra of SN 2009ip around the time of the first bump. The [Ca {\sc ii}] beings to emerge at this epoch, which may indicate a lower density, but it is difficult to envisage a decrease in density causing a rise in luminosity.

\subsection{Is there Ni in SN 2009ip?}
\label{s4c}

The final epochs of photometry obtained for SN 2009ip before it disappeared behind the Sun appear to show a flattening in the redder bands. Given that the lightcurve displayed undulations prior to this (as discussed in Section \ref{s4b}), it is entirely possible that the apparent flattening is merely the start of a third bump in the lightcurve, rather than the beginning of any $^{56}$Ni-powered tail phase. We note that the overall decline in the lightcurve from peak (4.5 mag over 80 days in $V$) is much faster than that expected from the decay of $^{56}$Co (0.98 mag per hundred days).

From our final photometric point on December 29, we can estimate an upper limit to the mass of $^{56}$Ni synthesized in SN 2009ip (assuming it was indeed a genuine SN). We measure a bolometric luminosity of log(L)= 40.8 erg s$^{-1}$ for SN 2009ip on December 29. Using eq. 2 of \cite{Ham03}, and assuming that SN 2009ip exploded on 2012 Aug 1, we calculate a $^{56}$Ni mass of $<$0.02 \msol. This is an upper limit to the mass of $^{56}$Ni in SN 2009ip, as it is based on the earliest plausible date for a SN explosion \citep{Mau13}. If the 2012a was not a SN, but rather the star exploded at the start of the 2012b event, then the $^{56}$Ni mass must be less than 0.01\msol. Furthermore, \cite{Ham03} assumes that the sole energy source powering the luminosity of the tail is radioactive decay, whereas in the case of SN 2009ip there is at least some luminosity arising from interaction (as evidenced by the persistence of narrow emission lines until this epoch).

One caveat to this calculation is that the formation of dust can serve to mask the late time luminosity of SNe at optical wavelengths, and cause their $^{56}$Ni mass to be underestimated. However, in the case of SN 2009ip, we see no evidence for dust formation in either spectroscopy (no apparent blue shift of lines) or in photometry (no strong NIR excess). A second caveat is that we assume that $\gamma$-rays from the decay of $^{56}$Co are fully trapped within the ejecta. For the well-observed SN 1987A \citep{Sun90}, approximately 2/3 of $\gamma$-rays were still trapped at 500 d post-explosion. The ratio of the trapping timescale in SN 2009ip to that in SN 1987A will scale as $\left(M_{09ip}/M_{87A}\right)^{0.5}\left(v_{09ip}/v_{87A}\right)^{-1}$. If we assume that the velocities are roughly similar in the two events, and that the ejecta mass of SN 1987A was 12\msol, then this scaling implies that for $\gamma$-ray leakage to be significant at 100 or 150 d (depending on assumed explosion epoch), the ejecta mass in SN 2009ip should be less than 0.5 or 1.1 \msol\ respectively. Such a small ejecta mass is much lower than that typically seen in SNe, and appears inconsistent with the core-collapse of a massive star. Even with these caveats, the restrictive upper limit on the mass of any $^{56}$Ni in SN 2009ip appears robust.

\subsection{The energetics of SN 2009ip}
\label{s4d}

Integrating under the uvoir bolometric lightcurves of the 2012a and 2012b events, we find an observed radiated energy of 7.8$\times10^{47}$ ergs for the former and 1.8$\times10^{49}$ ergs for the latter. In a normal CCSN, the radiated energy over the first one hundred days is typically comparable to that radiated in the 2012b event, $\sim10^{49}$ ergs.

While the large energy of the 2012b event is suggestive of a genuine SN explosion, with core-collapse as the energy source, it is interesting to consider the possibility that the high luminosity is in fact completely due to the efficient conversion of kinetic energy from colliding ejecta. From the broad absorptions in the Balmer lines we see material at velocities up to 12,000 \kms, however we adopt 7,500 as a more typical velocity for the bulk of the ejecta. If we assume 100 per cent efficiency in converting kinetic energy to radiation, then we only require 0.05 \msol of material at 7500 \kms to give rise to the luminosity of the 2012b event.

To check whether the assumption of high efficiencies in the conversion of kinetic energy is correct, we apply the Rankine--Hugoniot conditions to the shock. For ejecta moving at $v_{ej}$ hitting a reverse shock at $v_{rs}$, the efficiency $\epsilon$ in converting kinetic energy to luminosity is given by

\begin{equation}
\epsilon = \zeta {{ v_{ej}^2 - ( \eta   v_{ej} + (1-\eta) v_{rs})^2       }\over{v_{ej}^2    }}
\end{equation}

where $\zeta$ is the porosity ($Area_{clumps}/Area_{CSM}$) to account for clumpiness in the CSM, and $\eta$ is the ratio of post- and pre-shock velocities in the rest frame of the shock. $\eta$ depends on the equation of state of the gas, ranging between 1/4 and 1/7 for a perfect gas and a radiation dominated gas respectively. If we take the reverse shock to be moving at approximately the velocity of the photosphere derived from our blackbody modelling, 1200 \kms, and the ejecta to be moving at a conservative estimate of 7500 \kms, then we derive efficiencies of around 90 per cent. These efficiencies will be reduced by $\zeta$, although even if 50 percent of the CSM shell was clumps and 50 percent were holes, then we would still have $\epsilon\sim0.45$. In this case, we would still only require $\sim$0.1 \msol of ejecta to power the interaction luminosity.

\section{Modelling the CSM around SN 2009ip}
\label{s5}

As a first attempt at understanding the physical conditions in SN 2009ip, we have constructed a spherically symmetric CSM model of SN 2009ip with a density profile R$^{-2}$ as detailed in Appendix \ref{sa1}. While the real structure of SN 2009ip is undoubtedly much more complex, with multi dimensional effects and probably multiple shocked regions \cite{Dwa12}, less complex models can hopefully be used to derive order of magnitude results.

\begin{figure}
\includegraphics[width=0.5\textwidth]{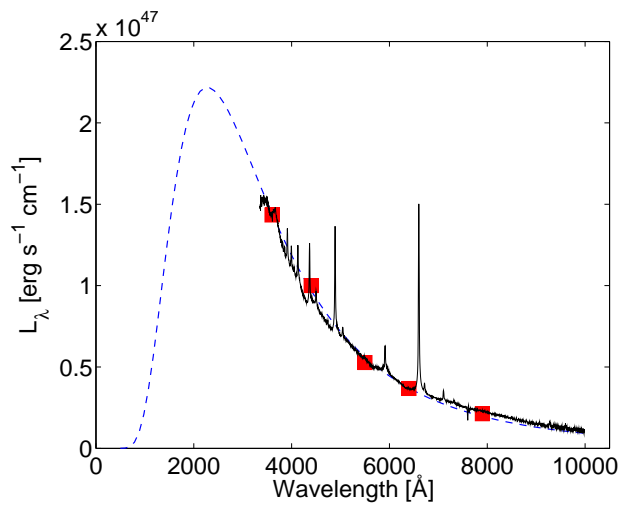}
\caption{The (dereddened) spectrum at Oct 8 2012 (black, solid), together with (dereddened) photometry (red points) and black-body fit to the photometry (blue, dashed). The derived temperature is 12650 K.}
\label{fig:bbfit}
\end{figure}

\begin{figure}
\includegraphics[angle=270,width=0.45\textwidth]{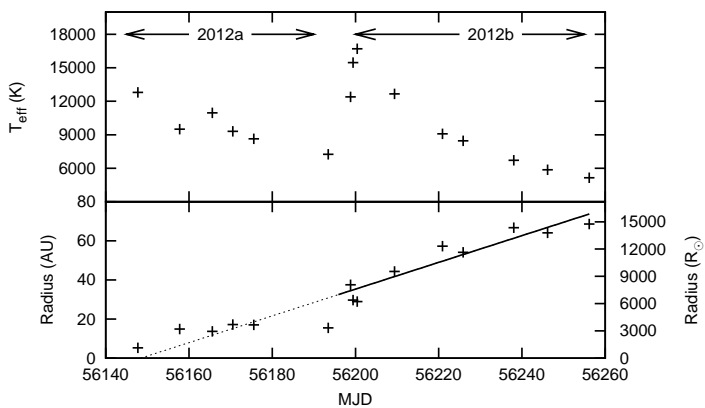}
\caption{{\it Upper panel}: The temperature evolution of SN 2009ip as measured from black-body fits to optical photometry at selected epochs. The approximate dates of the 2012a and 2012b events are indicated. {\it Lower panel}: The photospheric radius of SN 2009ip as determined from blackbody fits to optical photometry. The solid line is a least squares fit to the radii measured for the 2012b event only, while the dashed line is the extrapolation of that fit to the earlier 2012a event. The slope of the fit is 1200 \kms.}
\label{fig:t_r}
\end{figure}

We perform a least-squares fit for black-body radii and temperatures ($4\pi R_*^2 \pi B_\lambda(T_*) = L_\lambda$) to the observed, de-reddened, $UBVRI$ photometry for selected epochs (typically the ones with full photometric coverage). The fit for the epoch we model in detail in Section \ref{s5} (Oct 8 2012) is shown in Fig. \ref{fig:bbfit} as an example, while the time evolution of the derived radius and temperature is shown in Fig. \ref{fig:t_r}. The radius at the start of the 2012b event derived from a black body fit is also  consistent with the maximum radius derived from light travel time considerations using the high cadence photometry presented by \cite{Pri13}. The evolution observed over the 2012b event is quite similar to that seen in SN 1994W, which peaked at 15000 K and declined to 7000 K over 70 d, but hotter than that of SN 2009kn, which only peaked at 10000 K \citep{Kan12}. However, as the first spectra of SN 2009kn were obtained later than for SN 1994W, it is possible that the maximum temperature was in fact higher. Comparing to the radius evolution of SNe 1994W and 2009kn, we see that the initial photospheric radii of both these objects (R$\sim$8,500\rsun) was larger than that of SN 2009ip during the 2012a event (R$\sim$3000\rsun), but that they expanded with similar velocities to the 1200 \kms\ observed in the 2012b event ($v_\mathrm{SN1994W}=$1000 \kms, $v_\mathrm{SN1994W}=$1000 \kms; $v_\mathrm{SN2009kn}=$600 \kms).

We use the photospheric temperatures and radii derived as radiative boundary conditions for computing the ionization and excitation balance of hydrogen in the CSM as described in Appendix \ref{sa1}. We do not solve for temperature, but assume $T = T_*$, which should be a reasonable approximation. 

In particular, we investigate the first epoch after maximum light when we have an optical spectrum, MJD = 56209 days, when $T_* \approx 12,600$ K and $R_* \approx 44$ AU. The lack of any strong broad lines from the ejecta at this epoch, combined with the large photospheric radius is possibly due to the formation of a cool dense shell at the interface between the outer ejecta and CSM, which becomes optically thick in the Paschen continuum for a few weeks if the CSM density is high enough \citep{Chu01}. Outside this will be a region of shocked CSM, and outside the forward shock a region of unshocked CSM, radiating due to ionization by the photospheric radiation field (as well as from X-rays from the forward shock, which we however do not include here). Another possible explanation for the lack of broad absorptions is that the fast material is efficiently ionized by the emission from the shock region.

We explored the sensitivity of the model to having a region between the photosphere and the CSM, but do not find a significant difference.

For the inner parts of the CSM, the photospheric radiation field is strong enough that radiative processes are important not only for the ionization balance but also for the NLTE solutions to the excitation balance. 

We may therefore not use standard recombination theory, which assumes that the external radiation field is negleglible for the statistical equilibrium (as it is in normal HII regions where the dilution factor $(R/R_*)^2$ is large).

We explore models with different assumptions for the velocity gradient in the CSM (as discussed in Appendix \ref{sa1}). Fig. \ref{fig:LHa} shows that a model with a zero velocity gradient (i.e. a constant velocity flow throughout the emitting region) is not successful in reproducing the observed strength of H$\alpha$ \footnote{The observed value was obtained by measuring the relative amount of flux in the line compared to the total flux in the spectrum (3300 - 10000 A), plus estimated fluxes outside the spectral range from the black-body fit. This gives a fraction of $5\times10^{-3}$. Assuming no flux outside the spectrum, the value is 0.012.}. This is a consequence of the efficient line trapping in a constant velocity flow; when the density becomes high enough to trap any significant fraction of the photospheric radiation ($n \gtrsim 10^9$ cm$^{-3}$), H$\alpha$ has such a high optical depth that $n=3$ (the level creating H$\alpha$) becomes depopulated by collisions and photoionization rather than spontaneous decay. 

\begin{figure}
\centering
\includegraphics[width=1\linewidth]{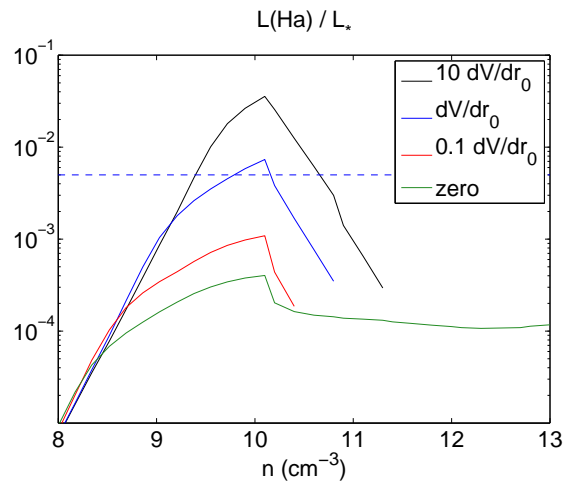}
\caption{Ratio of  H$\alpha$ and photospheric luminosity, on Oct 8 2012 (first spectrum after maximum light). The observed value ($3\times10^{-3}$) is marked with a dashed line. The models with a velocity-gradient are terminated at the density where the Sobolev limit is no longer reached (at high densities they behave as the model with zero gradient).}
\label{fig:LHa}
\end{figure}

\begin{figure*}
\subfigure[H$\alpha$/H$\beta$ ratios as a function of density in our models for the CSM lines around the maximum of the 2012b event. The observed value is indicated with a dashed line.]{
\includegraphics[width=0.47\textwidth,angle=0]{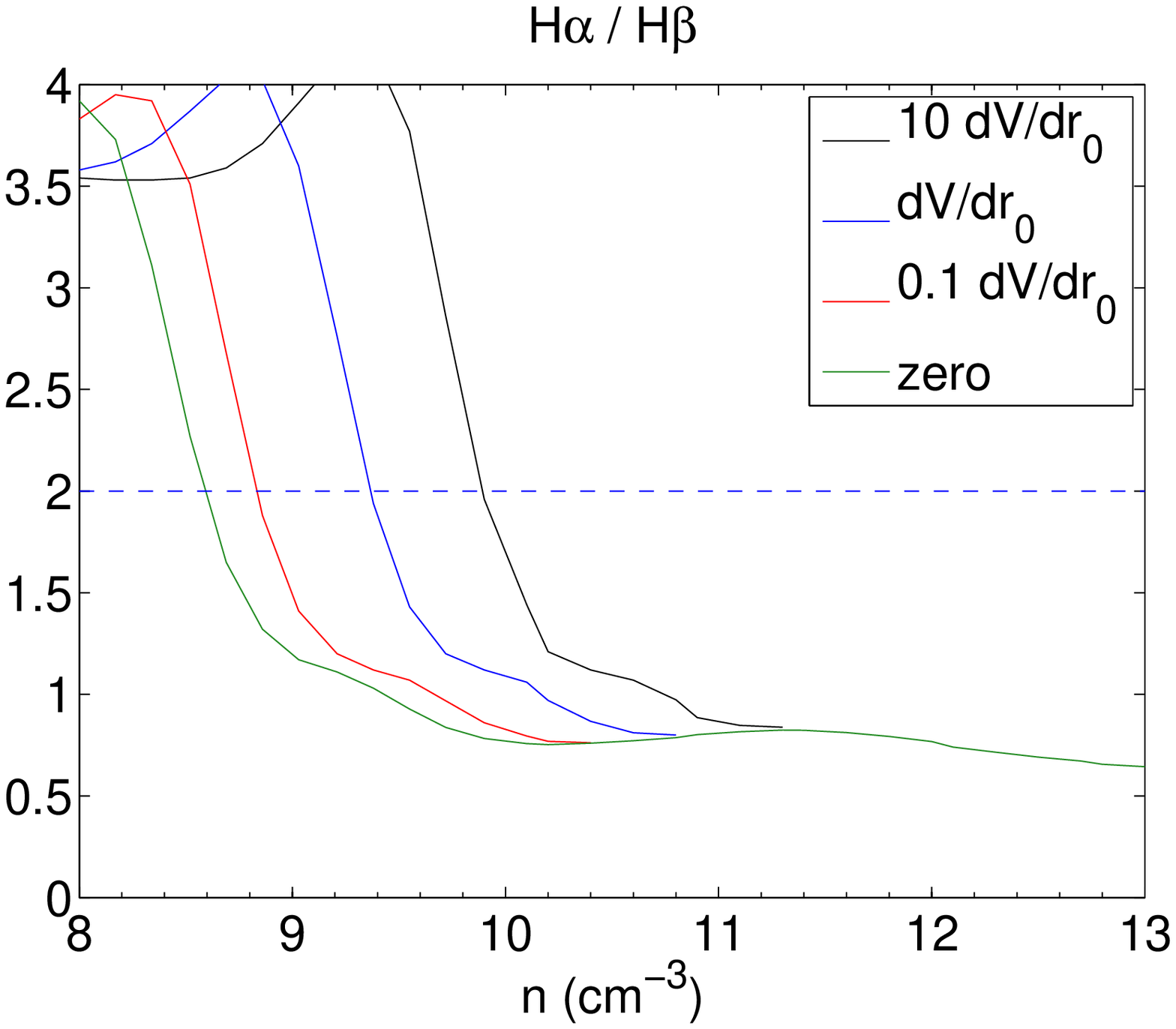}
\label{fig:balmerdec}
}
\subfigure[Same as Fig. \ref{fig:balmerdec}, but for the H$\gamma$/H$\beta$  ratio.]{
\includegraphics[width=0.47\textwidth,angle=0]{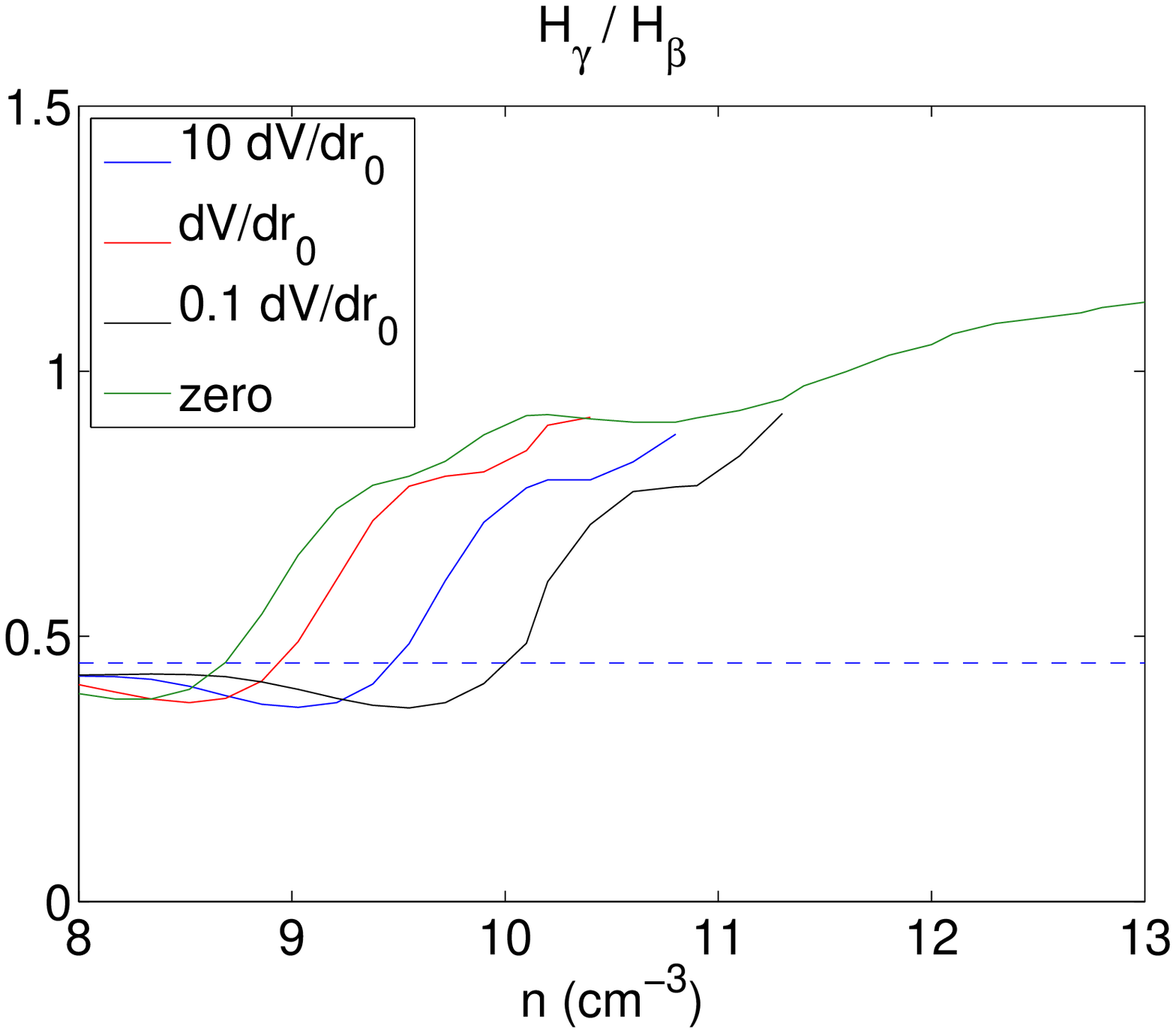}
\label{fig:hgamma}
}
\subfigure[Model H$\alpha$ (solid) and H$\beta$ (dashed) optical depths around the peak of the 2012b event.]{
\includegraphics[width=0.47\textwidth,angle=0]{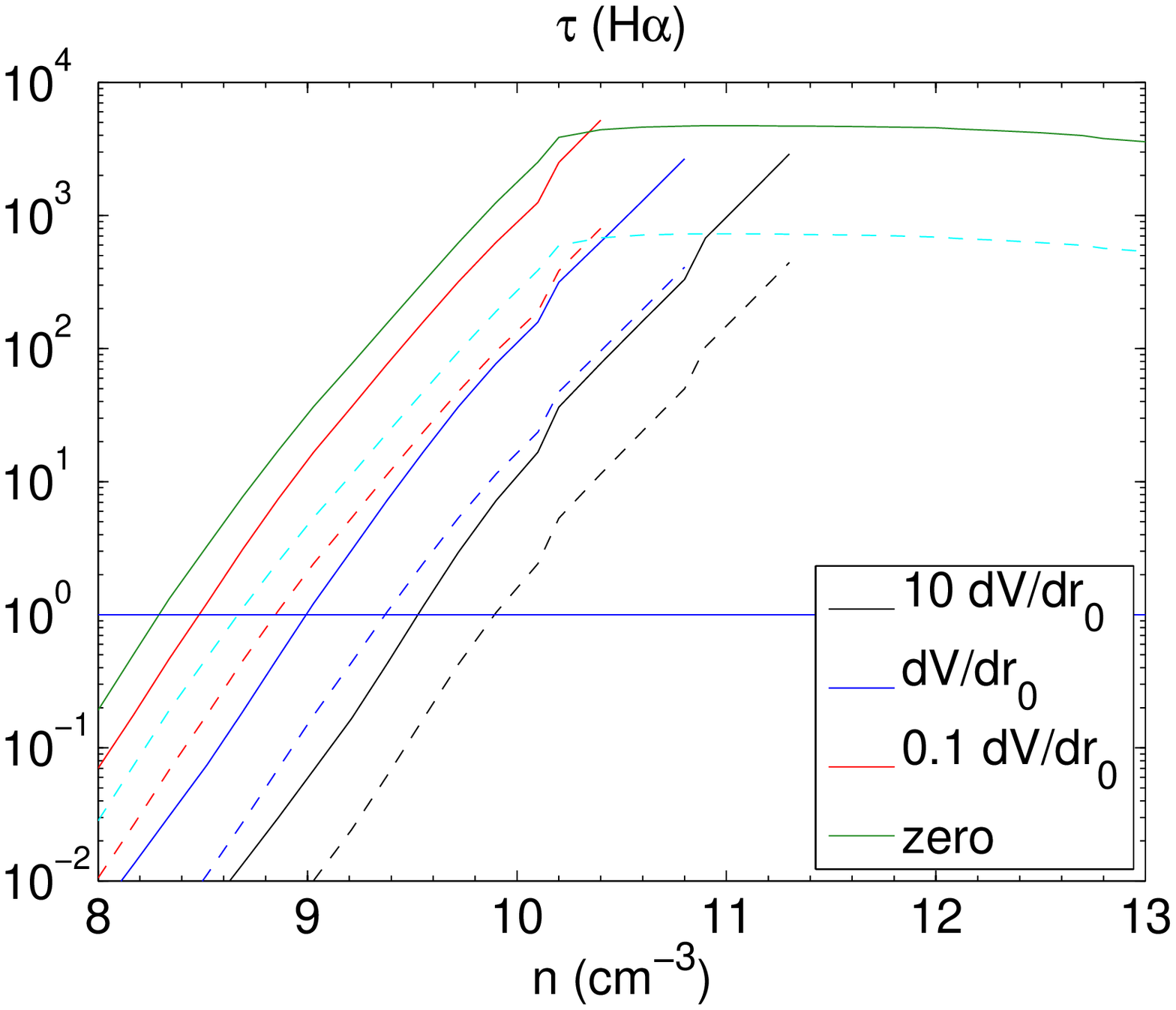}
\label{fig:balmerdepths}
}
\subfigure[Electron scattering optical depths in our models for the CSM lines around the peak of the 2012b event.]{
\includegraphics[width=0.47\textwidth,angle=0]{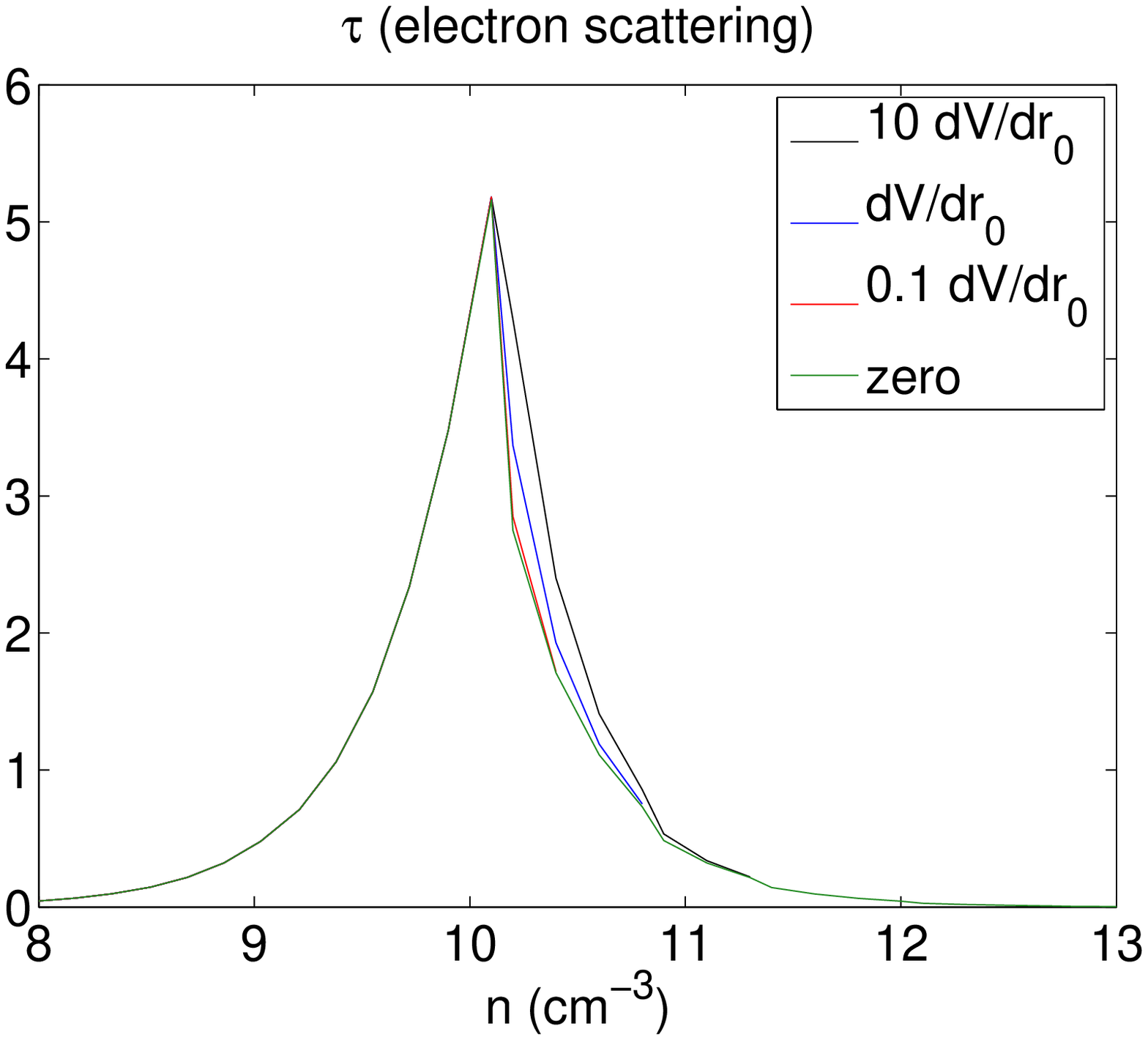}
\label{fig:taues}
}
\caption{Results from our modeling as described in Section \ref{s5}.}
\label{fig:modelling}
\end {figure*}

Models with velocity gradients $dV/dr \gtrsim dV/dr_0$ ($\sim1.4\times10^{-7}$; our estimated velocity gradient, found by dividing the CSM velocity by the scale size of the region), produce more promising values of the H$\alpha$ luminosity for densities $n \sim 10^{9}-10^{10}$ cm$^{-3}$. Turning to Fig. \ref{fig:balmerdec} and \ref{fig:hgamma}, we see that these models also produce correct line ratios of $L(Ha)/L(Hb) \approx 2$ and $L(H\gamma)/L(H\beta) \approx 0.45$. As Fig.  \ref{fig:balmerdepths} shows that H$\alpha$ is marginally optically thick in these models, the reason for the damped Balmer decrement seems to be that H$\alpha$ has an  optical depth of a few, but H$\beta$ is optically thin, a scenario discussed by \cite{Lev12}.

Fig. \ref{fig:taues} shows that all models predict electron scattering optical depths of order unity for densities between $n=10^9 - 10^{11}$ cm$^{-3}$. Since the Balmer lines show distinct broad wings that likely arise due to electron scattering with optical depth of order a few \citep{Chu01, Chugai2004,Des09,Smi10b}, this further strengthens the idea that the CSM density lies in this interval.

As made clear by the previous figures, the high density $n \gtrsim 10^{13}$ cm$^{-3}$ proposed by Levesque as one possibility for the low Balmer decrement is not succesful in reproducing the observables. At these high densities the Balmer lines have high optical depth for any reasonable assumption about the velocity field, and as the level populations also approach LTE, the line ratios approach their blackbody values. As an example, the $H\gamma/H\beta$ ratio goes towards 1.3 at high densities (and $T=12000$ K), whereas the observed value is $\sim 0.45$, as reproduced by the NLTE models around $n=10^{9}-10^{10}$ cm$^{-3}$. A high-density model would need $T\sim 2500$ K to reproduce $H\gamma/H\beta$ ratio, but at that temperature the Balmer decrement has value 4.8, much too high. The other model result that clearly rules out a $n \gtrsim 10^{13}$ cm$^{-3}$ scenario is that that the electron scattering optical depth would be much smaller than unity. 

The number density for a constant mass loss rate at a radius $r$ is
\begin{equation}
\begin{split}
n(r) = \frac{\dot{M}}{\bar{A}m_p 4\pi r^2 v(r)} = 5\cdot 10^{10} \mbox{cm}^{-3} \left(\frac{\dot{M}}{1~M_\odot~\mbox{yr}^{-1}}\right)\times \\
	\left(\frac{r}{40~\mbox{AU}}\right)^{-2}\left( \frac{V}{10^3~\mbox{km}~\mbox{s}^{-1}}\right)^{-1}
\label{eq1}
\end{split}
\end{equation}
where $\dot{M}$ is the mass-loss rate, $\bar{A}$ is the mean atomic weight (1.6 for a normal H/He mix), $m_p$ is the proton mass, and $v(r)$ is the wind velocity. The derived density of $n \sim 10^{9}-10^{10}$ therefore suggests as mass-loss rate of $\dot{M} \sim 0.01-0.1 M_\odot$/year. We caution however, that this is likely only correct to an order of magnitude, due to the uncertain assumption of a constant mass-loss rate.

Using more sophisticated modelling, \cite{Des09} find a similar Balmer decrement of $\sim$2 for SN 1994W from a model with a density of $10^9-10^{10}$ cm$^{-3}$ and with a velocity of $\sim$800\kms, \cite{Chugai2004} also require a similar density in order to reproduce the electron scattering wings. That these values are in close agreement with our simpler calculations lends credence to our approach.

\section{Analysis}
\label{s6}

There are thee questions which are key to understanding SN 2009ip. Firstly, what is the significance of the surprisingly remote location of SN 200ip; secondly, were either of the 2012a/b events a SN explosion or a mega-outburst of an impostor; and thirdly, what powers the luminosity of the 2012a/b events? As these questions are linked, we will address them in turn.

\subsection{The lonely fate of SN 2009ip}
\label{s6a}

\begin{figure}
\includegraphics[angle=0,width=0.5\textwidth]{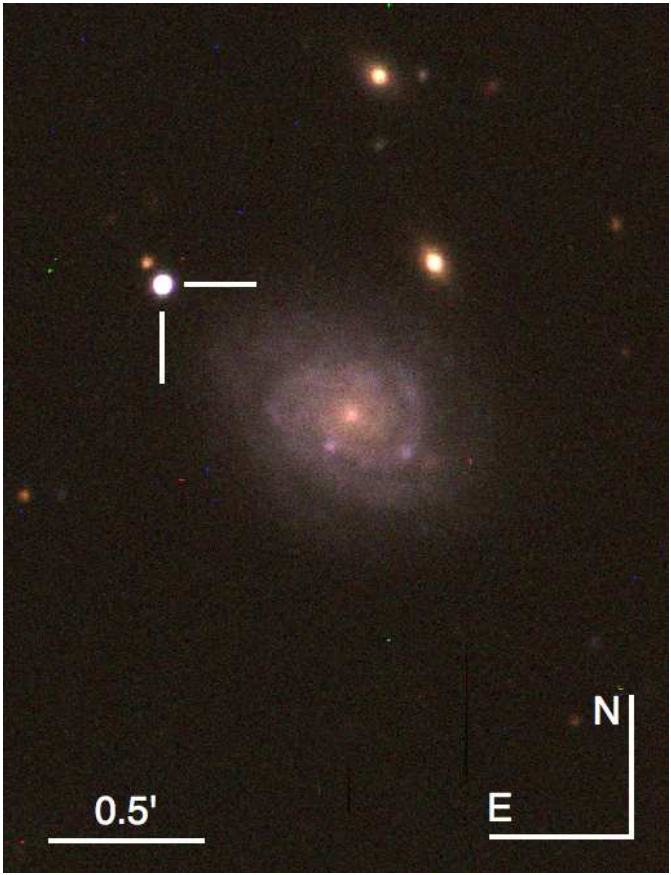}
\caption{{\it BVr} color composite of SN 2009ip (marked with cross hairs) and NGC 7259 obtained with the Liverpool Telescope + RATCAM on 2012 November 13. Scale and orientation are as indicated. The 0.5\arcmin\ scale bar corresponds to a physical distance of approximately 3 kpc at the distance of NGC 7259.}
\label{fig:finder}
\end{figure}

One of the most striking features of SN 2009ip is its isolated location, as illustrated in Fig. \ref{fig:finder}. SN 2009ip is at a projected distance of $\sim$0.75\arcmin\ from the center of NGC 7259, placing it at a deprojected physical distance of of 4.3 kpc. To measure the radial extent of the galaxy, we took the deepest $R$ band images of SN 2009ip taken with EFOSC2 under 1\arcsec\ seeing or better (from 2011 June 23), and stacked them to create a deep 1200s exposure. No faint spiral arm or other obvious structure can be seen at the SN location, in fact the nearest extended source to SN 2009ip in NGC 7259 is $\sim$10 \arcsec\ distant. We measure a surface brightness at the SN location of R=25 mag arcsec$^2$. The radial extent of NGC 7259 has been measured at 0.6' to 0.78' for surface brightness of 25 and 27 mag arcsec$^{2}$ respectively (as listed in NED, from \citealp{Lau89}). Hence although there is little star formation in the vicinity of SN 2009ip, it appears to reside within the optical disk of the galaxy. 

\begin{figure}
\includegraphics[angle=270,width=0.5\textwidth]{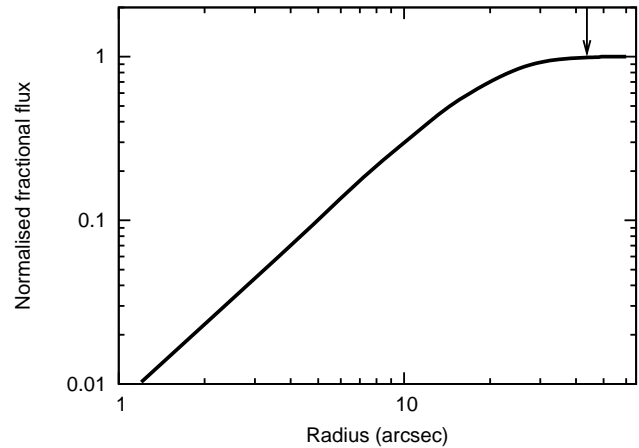}
\caption{Enclosed $R$ flux from NGC 7259 as a function of radius, normalized to total flux. The arrow indicates the position of SN 2009ip, which lies outside the detectable extent of the galaxy.}
\label{fig:surface}
\end{figure}

After masking the nearby galaxy 2MASXi J2223042-285647 in our deep image, we performed aperture photometry on NGC 7259 using circular apertures between 5 and 250 pixels (i.e. to a maximum radius of 1\arcmin). The sky background was measured in a blank region of the image  several arc minutes distant from NGC 7259 to avoid any unintended contamination. We plot the enclosed flux from NGC 7259 as a function of radius in Fig. \ref{fig:surface} in a similar fashion to \cite{And09}. The fraction of {\it R} band light within the position of SN 2009ip, ${F_r}_R$, is $\simeq$1. \citeauthor{And09} found that the mean ${F_r}_R$ for Type IIn SNe and SN impostors were slightly larger than for Type IIP SNe, and that there was a deficit of both impostors and Type IIn SNe in the central regions of galaxies. While SN 2009ip is in agreement with this trend, it is an extreme example, and indeed has a larger value of ${F_r}_R$ than any of the interacting transients in their sample. This is also in accord with \cite{And12}, who found that SN impostors and Type IIn SNe are the least correlated with ongoing star formation as measured from H$\alpha$ and {\it UV} imaging. It is worth noting, however, that the ``SN impostor'' category of \cite{And12} is extremely heterogeneous, and contains quite dissimilar transients (for example SN 2008S, \citealp{Bot09} and SN 2002kg, \citealp{Mau06,Wei05}). Nonetheless, the location of SN 2009ip is quite unique among both Type IIn SNe, and SN impostors. \cite{And12} have used the argument that the lack of correlation of Type IIn SNe with H$\alpha$ emission implies they have lower mass progenitors than the bulk of the CCSN population. While SN 2009ip is certainly isolated from any strongly starforming regions, it has (without doubt) a high mass progenitor star. 

The HST+WFPC2 $F606W$ images presented by \cite{Fol11} are also of use for constraining the environment of SN 2009ip. Using {\sc hstphot} \citep{Dol00}, we detect only a single source\footnote{We are excluding here the foreground star (R$\sim$18.4) to the NE of SN 2009ip.} within a 4\arcsec\ radius of SN 2009ip. We calculate a limiting magnitude of {\it F606W}$>$25 mag, implying that there is only a single source with a $R$ magnitude brighter than -6.6 within a 500 pc radius of SN 2009ip.

The apparently isolated position of SN 2009ip is of interest for two reasons. Firstly, as a LBV with a ZAMS mass of $\sim$50-60 \msol, the progenitor of SN 2009ip could be an example of isolated starformation. While there are examples of isolated massive stars in the LMC (for example, VFTS 682, \citealp{Bes11}), these are still within a few tens of parsec of a massive star forming region, and in the case of VFTS 682 have been suggested to have been dynamically ejected from a young cluster with a top-heavy IMF \citep{Ban12}. For SN 2009ip, there is no obvious nearby cluster, even with a runaway velocity of 50\kms, permitting it to cover a distance of 500 pc in 10$^7$ years. Secondly, the extreme radial location also points towards a sub-solar metallicity. Applying the metallicity gradient of \cite{Boi09} to the position of SN 2009ip, and estimating the characteristic metallicity of NGC 7259 from its {\it B} band magnitude, we find a local metallicity of 12+log(O/H) = 8.5 dex. This value is significantly lower than the locus of SN metalliciites derived by \citeauthor{Boi09} using the same technique on a sample of SN host galaxies. Unfortunately there is no nearby H {\sc ii} region where we could obtain a spectroscopic estimate of the metallicity, while the auroral [O {\sc ii}] and [O {\sc iii}] lines are not detected in our spectra precluding a direct measurement of the electron temperature and metallicity.

\subsection{Has SN 2009ip exploded as a SN?}
\label{s6b}

Ultimately, the most definitive proof that a star has exploded as a CCSN is the disappearance of its progenitor in late time images taken after the explosion has faded. While it should be possible to eventually test this for SN 2009ip, at the time of the final photometry obtained of SN 2009ip at the end of December, the transient was still $\sim$4.5 mag brighter than the progenitor detected by \cite{Fol11}. If the 2012b event continues its decline rate from maximum of $\sim$3.5 mag in 70 d, then by the time SN 2009ip reappears in mid-April 2013, we may expect it to be fainter than the magnitude of the progenitor if it was a genuine SN.

Another critical piece of evidence for a SN explosion is the detection of nucleosynthesized material in the ejecta. This may be either explosively produced isotopes (e.g. $^{56}$Ni) or elements synthesized during the latter stages of stellar evolution (e.g. $^{16}$O). In SN 2009ip we do not see any evidence for freshly synthesized material; all the lines present in the Dec 20 spectra were also present in September 2009. We also see none of the usual lines which are typically seen in the nebular spectra of CCSNe such as [O {\sc i}] 6300,6364 \AA, Mg {\sc i}] 4571 \AA\ and Mg {\sc i} 1.5$\upmu$m. Against this however, we must note that many Type IIn SNe do not show these lines until quite late phases, due to the low gas densities of $\sim$10$^9$ cm$^{-3}$ required. Given the precipitous decline of SN 2009ip from the 2012b peak, one may expect a reasonably prompt transition to the nebular phase.

Another signature of a CCSN is the presence of a radioactively powered tail phase, which at late times follows the decay rate of $^{56}$Co. From our last photometric point, we set a limit of $<$0.02 \msol\ of Ni in SN 2009ip. Such a low ejected Ni mass is hard to reconcile with the core-collapse of a very massive star, although a similarly low value of $\lessapprox$0.023 \msol\ was set for SN 2009kn \citep{Kan12}, albeit with the caveat that the tail may not be Ni-powered. \cite{Ume08} find that models with a ZAMS mass of 60-80 \msol\ produce between 0.8 and 7.9 \msol\ of Ni, depending on explosion energy. The low upper limit to the ejected Ni mass could be explained if much of the Ni did not escape, but rather underwent fallback onto the compact remnant. Such a scenario was proposed for SN 1997D \citep{Tur98}, although in the case of fallback the ejecta velocities are expected to be closer to 10$^3$ than 10$^4$ \kms.

We note that a low ejected Ni mass was also derived for the Type IIn SN 1994W \citep{Sol98}. In this case, \cite{Chugai2004} claimed that the transient arose from the interaction of a SN with the hydrogen envelope of the star, which had been ejected 1.5 years prior. \citeauthor{Chugai2004} proposed that the ejection of the envelope could be due to a Ne burning flash in the O-Ne-Mg core of an 11 \msol\ star; while the low progenitor mass would also account for the small ejected Ni mass. The explosion of a Super Asymptotic Giant Branch (SAGB) star was also claimed for SN 2011ht by \cite{Mau12}, and suggested as a possibility by \cite{Kan12} for SN 2009kn. In the case of SN 2009ip however, the massive progenitor is clearly inconsistent with such a scenario. An alternative model for SN 1994W was proposed by \cite{Des09}, who suggested that the collision of two ejected shells with an efficient conversion of kinetic energy to luminosity could also power the transient. In the colliding shells model, the progenitor star is not necessarily destroyed, although the lack of pre-explosion imaging coupled with its crowded environment makes this difficult to test for SN 1994W with deep, late-time imaging.

Less compelling indicators of a CCSN include high velocities and luminosities. In the case of the former, the observation by P13 of material at velocities of $\sim10^4$ \kms\ in 2011, long before the start of the 2012a and 2012b events, suggests the material can be accelerated to high velocities {\it without} a core-collapse explosion. The bright absolute magnitude of SN 2009ip (V=-17.7) is certainly suggestive of core collapse, however, as we have shown in Section \ref{s4d} the kinetic energy of a relatively small amount of material at the velocities observed can power the lightcurve. Furthermore, some of the most luminous SNe known have been suggested to arise from outbursts which do not (at least immediately) destroy the progenitor star \citep{Woo07}.

From the available evidence -- and remaining mindful of the truism that ``absence of evidence is not evidence of absence'' -- we suggest that thus far there is no evidence that SN 2009ip did indeed undergo core-collapse to produce either the 2012a or 2012b events.

\subsection{What powers the luminosity of the 2012a and 2012b events?}
\label{s6c}

Any scenario for SN 2009ip needs to explain the following key characteristics:

\begin{itemize}
\item An intrinsically luminous, a most likely very massive progenitor star, probably with a low metallicity.
\item Multiple eruptions over at least a three year period, which eject material at velocities of up to 10$^4$ \kms.
\item A brief, 40 day eruption. 
\item A subsequent more luminous eruption, with a fast rise time, which decays rapidly from peak.
\item An ejected $^{56}$Ni mass $<$0.02 \msol.
\item No observed synthesized material, either from explosive nucleosynthesis or late-stage stellar evolution. 
\end{itemize}

There have been several suggestions as to the energy and nature of the 2012a and 2012b events. \cite{Mau13} argued that the 2012a event was the explosion of SN 2009ip as a weak CCSN with a low ejected Ni mass, while the 2012b event was caused by the ejecta hitting the CSM. P13 further suggested that if SN 2009ip was indeed a genuine SN, then the 2012b event may be the explosion of SN 2009ip, while the 2012a event was an outburst. However, P13 also advanced an alternative scenario that SN 2009ip has not exploded as a SN, but rather that the lightcurve and spectra could be explained by the collision of two massive shells of material ejected by the progenitor \citep{Des09}. Such an eruption could be powered by the pulsational pair instability \citep{Woo07}. A final model for SN 2009ip was proposed by \cite{Sok13}, who favoured a merger of a massive LBV with a binary companion, in a ``mergerburst'' model. \citeauthor{Sok13} draw comparison to the enigmatic transient V838 Mon, for which a similar scenario was proposed.

\begin{figure}
\includegraphics[angle=270,width=0.5\textwidth]{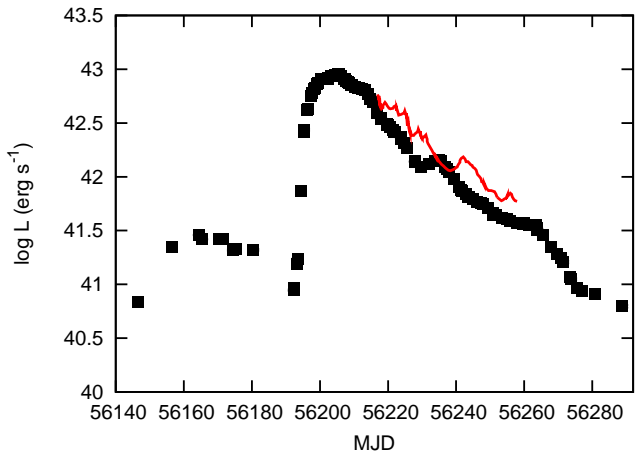}
\caption{Comparison of the bolometric lightcurve of SN 2009ip to the model prediction of \protect\cite{Sok13}.}
\label{fig:sok}
\end{figure}

\cite{Sok13} predicted a lightcurve evolution for SN 2009ip from the peak of the 2012b event (the last epoch where published data was available to them) as shown in Fig. \ref{fig:sok}. The overall match to the lightcurve of SN 2009ip is striking; the decline rate is similar, as is the amplitude of the undulations. However, \citeauthor{Sok13} also predict that only a small fraction of the ejecta should have velocities in excess of 5000 \kms, while in SN 2009ip we see the deepest absorption troughs at higher velocities than that, over three months after the 2012a peak.

As discussed in Section \ref{s6b}, while a weak CCSN explosion in the 2012a event cannot be entirely ruled out, we see no conclusive evidence in favor of it. Furthermore, while a weak explosion could explain the low $^{56}$Ni mass, we would not expect to observe the high ejecta velocities seen. If the 2012b event was the SN, then the upper limit to the $^{56}$Ni mass is even more restrictive. Whether a $\sim$60 \msol\ star can explode, suffer sufficient fallback to hide almost all its Ni, but still retain a strong enough explosion to drive the ejecta outwards at upwards of 10$^4$ \kms\ remains to be seen.

In the canonical model for the Type IIn SN, the luminosity is powered by interaction between ejecta and the CSM. At the start of the 2012b event, the radius of the photosphere is around 30 AU, and is expanding with a velocity of 1200 \kms, as can be seen in Fig. \ref{fig:t_r}. If the observed photosphere is formed at the cool dense shell, then any ejecta moving with a velocity of $>$1000 \kms should reach this over the timescales of the 2012b event. Any such ejecta will pass through the shocks, being decelerated and heated in the process. The fact that ejecta with velocities close to 10$^4$ \kms\ is still observed in our last spectra indicates that either the photosphere is not at the interaction region - or that the cool dense shell/interaction region is clumpy, and that some of the fast ejecta can pass through it unhindered. The latter scenario could also help explain why even at the peak of the 2012b event there are weak broad absorptions.Such clumpiness is almost a certainty, given likely instabilities associated with the the multiple shocks passing through the CSM due to past explosive events.

An alternative arrangement to explain the simultaneous presence of high velocity absorption and interaction dominated line profiles could be the a disk or torus-like geometry, as suggested by \cite{Lev12}. In such a scenario, most of the CSM lies in a disk, the polar axis of which is inclined relative to our line of sight. A fraction of the fast ejecta will impact the disk and give rise to the interaction, depending on the thickness of the disk and the precise emitting geometry.

We suggest that the pulsational-pair instability models of \cite{Woo07} and \cite{Cha12} reproduce well the observed characteristics of SN 2009ip, and that the high luminosity seen during the 2012b event is probably powered by colliding shells. The 2012a event could also be powered by interaction, alternatively it could be simply due to the internal energy released during an outburst. Either way, this prediction can be tested by observing the surviving progenitor star (or finding it has disappeared) once SN 2009ip emerges from behind the Sun in April 2013. Accurate measurements of the metallicity at the site of SN 2009ip would also be of interest, given the expectation that pulsational-pair instability eruptions should only occur at low metallicity.

\section{Conclusions}
\label{s7}

We have presented intensive UV, optical and NIR spectroscopy and photometry of the interacting transient SN 2009ip. The transient was intrinsically luminous at the peak of the 2012b event, but over the following weeks faded much more rapidly than is typically seen in CCSNe. From our photometry we have set a robust upper limit of $<$0.02 \msun to the mass of any $^{56}$Ni synthesized in SN 2009ip. This limit can be considered secure with respect to both $\gamma$-ray leakage and dust formation. Spectoscopically, the transient shows a blue continuum with narrow emission lines from H and He, before developing strong broad absorption and emission in both H and He, along with Na and Ca. At no point do we see the nebular emission from synthesized material which is the hallmark of late stage evolution in CCSNe.

From our modelling of the CSM, we find densities of 10$^9$-10$^{10}$ are sufficient to explain the narrow Balmer lines and the observed Balmer decrement. We also find that a velocity gradient is required in the CSM to reproduce the H$\alpha$ to continuum ratio; such a gradient implies that the mass loss is eruptive rather than a steady wind.

We see no evidence for core-collapse in our data set, and in particular do not observe the appearance of nebular emission lines or detect a radioactive tail. We suggest that the properties of SN 2009ip can be explained with the efficient conversion of kinetic energy to optical luminosity via colliding shells, possibly resulting from a pair-instability driven outburst. 

A final question posed by SN 2009ip -- if it is indeed not a core-collapse SN -- is whether other interacting transients have been similarly misclassified. \cite{Des09} suggested that SN 1994W was not a CCSN, while the possibility was also raised by \cite{Kan12} for SN 2009kn. If the progenitor of SN 2009ip is seen to survive this latest eruption, it will confirm that a high peak luminosity and the detection of high velocity ejecta cannot be regarded as definitive proof of core-collapse in interacting transients.

\section{Acknowledgements}

This work is based on observations collected at the European Organisation for Astronomical Research in the Southern Hemisphere, Chile as part of PESSTO, (the Public ESO Spectroscopic Survey for Transient Objects Survey) ESO program 188.D-3003. 

We thank Erkki Kankare for useful advice and discussions on SN 2009kn.

Research leading to these results has received funding from the European Research Council under the European Union's Seventh Framework Programme (FP7/2007-2013)/ERC Grant agreement n$^{\rm o}$ [291222]  (PI : S. J. Smartt).

Parts of this research were conducted by the Australian Research Council Centre of Excellence for All-sky Astrophysics (CAASTRO), through project number CE110001020. J.P.A. acknowledges support by CONICYT through FONDECYT grant 3110142, and by the Millennium Center for Supernova Science (P10-064-F), with input from Fondo de Innovacin para la Competitividad, del Ministerio de Economa, Fomento y Turismo de Chile. G.P. acknowledges  support from the Iniciativa Cientifica Milenio grant P10-064-F (Millennium Center for Supernova Science), with input from ``Fondo de Innovaci\'{o}n para la Competitividad, del Ministerio de Econom\'{\i}a, Fomento y Turismo de Chile. K.T. acknowledges partial support by the Proyecto Gemini-Conicyt 32110024. F.B. acknowledges support from FONDECYT through Postdoctoral grant 3120227 and from the Millennium Center for Supernova Science through grant P10-064-F (funded by ÒPrograma Bicentenario de Ciencia y Tecnolog\'ia  de CONICYTÓ and ``Programa Iniciativa Cient\'ifica Milenio de MIDEPLAN''). F.E.B. acknowledges support from Basal-CATA (PFB-06/2007), CONICYT-Chile (FONDECYT 1101024), and Iniciativa Cientifica Milenio grant P10-064-F (Millennium Center for Supernova Science), with input from Fondo de Innovaci\'{o}n para la Competitividad, del Ministerio de Econom\'{\i}a, Fomento y Turismo de Chile. SB, MTB, AP, MT and PM are partially supported by the PRIN-INAF 2011 with the project "Transient Universe: from ESO Large to PESSTO". 

Observations were taken at the LBT as part of proposal INAF-2012B\_7. The Liverpool Telescope is operated on the island of La Palma by Liverpool John Moores University in the Spanish Observatorio del Roque de los Muchachos of the Instituto de Astrofisica de Canarias with financial support from the UK Science and Technology Facilities Council (Program ID OL12B38). Based on observations made with the Nordic Optical Telescope, operated on the island of La Palma jointly by Denmark, Finland, Iceland, Norway, and Sweden, in the Spanish Observatorio del Roque de los Muchachos of the Instituto de Astrofisica de Canarias.


\clearpage

\appendix

\section{NLTE modelling for the narrow H lines in SN 2009ip}
\label{sa1}

We compute NLTE solutions for the ionization and excitation balance of hydrogen, taking all radiative processes due to the photospheric radiation field into account, along with all bound-bound collisional transitions. We use a model atom including fine structure with $N=210$ $nl$-levels ($n=1-20$), with energy levels and A-values from NIST, bound-bound collision rates from \citet{Johnson1972} and \citet{Anderson2000}, photoionization cross sections from \citet{Verner1996} for $n=1$ and and scaled to $n\ge 2$ according to \citet{Rybicki}, and radiative recombination rates from \citet{Brocklehurst1971}. We do not include collisional ionization or its inverse process three-body recombination, which start to become important only at high densities ($n \gtrsim 10^{13}$ cm$^{-3}$).

We take radiative transfer effects into account by using escape probabilities. In the limit of small velocity gradients ($dV/dr \ll \Delta V_{thermal}/ \Delta r$) , we use the position-averaged escape probability in a uniform sphere \citep{OF}\footnote{Various flavors of escape probabilities may be used depending on the assumed geometry and density profile of the emitting region as well as the details of the radiative transfer approximations \citep{OF,Capriotti1965,Drake1980}, but for our purposes trying to select the ``best'' one is not meaningful, as they only differ by a factor of a few at most. Note that we also use the escape probability at line center, rather than the line-profiled average expression, as these
generally differ by less than a factor of two.}
\begin{eqnarray}
\label{eq:escapeprob}
\beta(\tau_{ij}) &=& \frac{3}{4\tau_{ij}}\left[1 - \frac{1}{2\tau_{ij}^2}+\left(\frac{1}{\tau_{ij}}+\frac{1}{2\tau_{ij}^2}\right)e^{-2\tau_{ij}}\right]\\
\tau_{ij} &=& \frac{A_{ij} \lambda_{ij}^3}{8\pi^{3/2}}\frac{g_j}{g_i}\left(n_i - n_j\frac{g_i}{g_j}\right) \left(\frac{\Delta V_{thermal}}{\Delta r}\right)^{-1}
\end{eqnarray}
where $i$ and $j$ refer to the levels ($i$ is the lower one here), $g_i$ and $g_j$ are the statistical weights, $A_{ij}$ is the Einstein coefficient for spontaneous decay, $\lambda_{ij}$ is the line wavelength, $\Delta V_{thermal}$ is the thermal velocity dispersion
\begin{equation}
\Delta V_{thermal} = \sqrt{\frac{2kT}{m_p}}
\end{equation}
and $\Delta r$ is the effective column length, given by
\begin{equation}
\Delta r = \frac{1}{n_i(R_{in}) - n_j(R_{in})\frac{g_i}{g_j}}\int_{R_{in}}^{R_{ion}} n_i(r) - n_j(r)\frac{g_i}{g_j} dr
\end{equation}
where $R_{ion}$ is the outer limit of the ionized region (computed iteratively).

In the limit of large velocity gradients ($dV/dr \gg \Delta V_{thermal}/ \Delta r$), we instead use the Sobolev approximation \citep{Sob60,Castor1970}
\begin{eqnarray}
\beta(\tau^s_{ij}) &=& \frac{1-e^{-\tau^s_{ij}}}{\tau^s_{ij}}\\
\tau^s_{ij} &=& \frac{A_{ij} \lambda_{ij}^3}{8\pi}\frac{g_j}{g_i}\left(n_i - n_j\frac{g_i}{g_j}\right) \left(\frac{dV(r)}{dr}\right)^{-1}
\end{eqnarray}

Since $\tau^s_{ij} / \tau_{ij} \approx \Delta V_{thermal}/\Delta V_{flow}$, where $\Delta V_{flow} \sim \frac{dV}{dr} \Delta r$, rapid flows ($\Delta V_{flow} > \Delta V_{thermal}$) have the effect of lowering the line optical depths, enhancing the escape probabilities.

A priori, the velocity gradients in the CSM of SN2009ip are unknown. The gradient could quite possibly be important, since typical expansion velocities of the narrow lines in the spectra are of order $\Delta V_{flow} \sim 10^3$ km s$^{-1}$, whereas the thermal velocities are of order $\Delta V_{thermal}\sim$ 10 km s$^{-1}$ (assuming $T \sim 10^4$ K). If the CSM was created by an eruption, the velocity field will be close to homology with a gradient of $\frac{dV}{dr} = t^{-1}$ where $t$ is the time since the eruption. If the CSM instead originates in a steady wind, and accelerations in the wind are small, the gradient can be insignificant over the emission region. 

We explore models with different assumptions about the velocity field. We introduce a default velocity gradient $\frac{dV}{dr}_0$ by taking the ratio of the typical CSM velocity ($10^3$ km s$^{-1}$) to the system scale size ($10^4~R_\odot)$, so 
\begin{equation}
\frac{dV}{dr}_0 = 1.4 \cdot 10^{-7}~~s^{-1}. 
\end{equation}

For an eruptive origin, this corresponds to $t=80$ days. We later try models with $\frac{dV}{dr} = \epsilon \frac{dV}{dr}_0$, for different values of $\epsilon$.

\subsection*{NLTE solutions}

The bound-bound radiative rates due to the external radiation field is computed by \citep[e.g.,][]{Castor1970, Elitzur1982}

\begin{equation}
R_{ij} = B_{ij} J_{\nu_{ij}} \beta(\tau_{ij})~s^{-1}
\end{equation} 
where $B_{ij}$ is the Einstein coefficient for photoabsorption ($i< j$) and stimulated emission ($i> j$), $J_{\nu_{ij}}$ is the mean intensity at the line frequency $\nu_{ij}$ (excluding any attenuation by the line), and $\beta(\tau_{ij})$ is the escape probability (described above). Close to the photosphere we have 
\begin{equation}
J_\nu = \frac{1}{4\pi} \int_0^{2\pi} d\phi \int_0^{\pi/2} B_\nu \sin \theta \cos \theta d\theta = \frac{1}{4} B_\nu
\end{equation} 
The photoionization rate for level $i$ is
\begin{equation}
\Gamma_i = \int_{\nu_0^{i}}^\infty \frac{4\pi J_\nu}{h\nu} \sigma_\nu^i d\nu~~~s^{-1}
\end{equation}
where $\sigma_\nu^i$ is the ionization cross section for level $i$, $\nu_0^{i}$ is the threshold frequency.
The total photoionization rate is
\begin{equation}
\Gamma = \sum_{i=1}^N \Gamma_i \frac{n_i}{n_{HI}}~~s^{-1}
\end{equation}
where $n_{HI} = \sum_{i=1}^N n_i$ and $n_i$ is the level population of level $i$ (computed iteratively). 

The ionization balance equation is, approximating the proton density to equal the electron density ($n_+ = n_e$), 
\begin{equation}
\Gamma n (1-x_e) = \left(\alpha + \alpha_{stim}\right) n^2 x_e^2
\label{eq:ion}
\end{equation}
where $x_e$ is the electron fraction $(=n_e/n$) $\alpha$ is the effective (including optical depth effects) spontaneous radiative recombination coefficient
\begin{equation}
\alpha = \sum_{i=1}^N \alpha_i \beta(\tau_i)
\end{equation}
where $\beta(\tau_i)$ is computed with the static escape probability formalism (Eq. \ref{eq:escapeprob}), and $\alpha_{stim}$ is the stimulated recombination coefficient

\begin{equation}
\alpha_{stim} = \sum_{i=1}^N \frac{g_i}{2}\left(\frac{2\pi m_e kT}{h^2}\right)^{-3/2}\int_{\nu_0^i}^\infty \frac{4\pi J_\nu}{h\nu} \sigma_\nu^i e^{-\left(h\nu - h\nu_0^i\right)/kT} d\nu
\end{equation}

Eq. \ref{eq:ion} is solved by the Newton-Raphson method. 

For a given solution to $x_e$ (and thereby for $n_+=n_e = nx_e$ and $n_{HI}=n\left(1-x_e\right)$), we then solve the set of statistical equilibrium equations for the excited levels of HI. For level $i$:
\begin{eqnarray}
\left[\alpha_i \beta(\tau_i) + \alpha_{stim}^i\right] n_e^2 + \sum_{j\ne i} n_j\left[C_{ji}n_e + A_{ji}\beta_{ji} + R_{ji}\right]\\
= n_i\left[\Gamma_i + \sum_{j\ne i} \left(A_{ij} \beta_{ij} + C_{ij}n_e + R_{ij}\right)\right]
\end{eqnarray}
where $\alpha_i$ is the specific radiative recombination coefficient, $C_{ij}$ are the bound-bound collision coefficients, and the other symbols are defined above. Since the escape probabilities $\beta_{ij}$ depend on the level populations themselves, the system is non-linear and is solved iteratively.

Together with the number conservation equation (we ignore contributions by He and metals to $n_e$, since this contribution is small unless very high ionization states are reached for the metals)
\begin{equation}
n_{HI} + n_e = n~
\end{equation}
we then iteratively solve for ionization balance, level populations, and line optical depths, for given values of $n$ and $T_*$ ($T$ is taken as = $T_*$).  This allows us to compute line optical depths and emissivity ratios for models in the Sobolev limit, where these quantities only depend on local conditions

If we assume a density profile for the CSM, we can also compute solutions for steady flows ($dV/dr = 0$), as well as integrated quantities such as total line luminosities and continuum optical depths in all the models.The size of the ionized region can be estimated by equating the number of absorbed continuum photons to the number of recombinations
\begin{equation}
\sum_{i=1}^N \int_{\nu_0^i}^{\nu_0^{i-1}} \frac{4\pi J_\nu}{h\nu}\left[1-e^{-\sum \tau_j(R_{ion})}\right] d\nu = \int_{R_{in}}^{R_{ion}} 4\pi r^2 n(r)^2 \alpha dr
\end{equation}
which can be solved for $R_{ion}$ given a density profile $n(r)$. The continuum optical depths are computed by Eq. \ref{eq:conttau}. 

For small enough densities, there is no solution as recombinations are too slow to bound an H {\sc ii} region, and the ionized region in then formally infinite (in actuality it is bounded by time travel limits, or when it encounters the interstellar medium.)

Given the solution to $R_{ion}$, the electron scattering optical depth is 
\begin{equation}
\tau_{es} = \sigma_{es}\int_{R_{in}}^{R_{ion}} n_e(r) dr~,
\end{equation}
which for $n_e(r) = n_e \left(R/R_{in}\right)^{-p}$ ($p \geq 2$) equals
\begin{equation}
\tau_{es} = \sigma_{es} n_e(R_{in}) R_{in} \frac{1 - \left(\frac{R_{in}}{R_{ion}}\right)^{p-1}}{p-1}~,
\end{equation}
where $\sigma_{es} = 6.65\e{-25}$ cm$^2$ g$^{-1}$. 
For the level populations, it is a reasonable approximation that $n_i(r) \propto r^{2p}$, since recombination rates scale with $n_e^2 \approx n^2$ (through the dilution of
the photospheric radiation field for photoexcitation and through the $n_e^2$ dependence for recombination. Thus, for the photoionization optical depths
\begin{equation}
\tau_i \approx \sigma_i(\nu_0^i) n_i(R_{in}) R_{in}\frac{1 - \left(\frac{R_{in}}{R_{ion}}\right)^{2p-1}}{2p-1}~.
\label{eq:conttau}
\end{equation}

To clarify, we thus compute the NLTE solutions explicitly for the density and radiation field specified at the inner region, and assume scalings of the solutions at larger radii according to $n_e \propto r^{-p}$ and $n_i \propto r^{-2p}$. In all the solutions presented in the paper, we use p=2, corresponding to a steady mass loss.

\section{Tables}

\onecolumn 
\begin{longtable}{llcccccr}
\caption{Optical photometry of SN 2009ip in the Landolt system, uncertainties are given in parentheses.}\\
\label{tab:opt}
Date            &       MJD             &       U                               &       B                               &       V                               &       R                               &       I                       &               Instrument      \\
\hline
09/10/12        &       56209.04        &       13.00           (0.04)  &       13.87   (0.02)  &       13.90   (0.01)       &       13.75   (0.01)  &       13.68   (0.01) &          EFOSC2                  \\
10/10/12        &    56210.27        &       13.04   (0.02)  &       13.93   (0.01)       &       13.93   (0.01)       &       13.78   (0.01)       &       13.69   (0.01)    &               EFOSC2                  \\
15/10/12        &       56215.24        &       13.15   (0.02)  &       14.01   (0.01)  &       13.97   (0.01)  &       13.79   (0.01)       &       13.72   (0.01)    &               EFOSC2          \\
16/10/12        &       56216.02        &       13.22   (0.01)  &       14.09   (0.01)  &       14.06   (0.01)       &       13.87   (0.01)       &       13.79   (0.01)    &               EFOSC2          \\
16/10/12        &    56216.71 &      13.34   (0.05)  &       14.16   (0.05)          &       14.03   (0.05)  &       -                               &       -                       &               UVOT            \\
16/10/12        &    56216.77 &      -                               &       14.28   (0.05)  &       -                               &       -                               &       -                       &               UVOT            \\
18/10/12        &       56218.01        &       -                               &       14.40   (0.02)  &       14.17   (0.02)       &       13.97   (0.02)       &       13.92   (0.04)    &               PROMPT          \\
19/10/12        &       56219.02        &       -                               &       -                               &       14.21   (0.01)       &       14.04   (0.03)       &       13.95   (0.03)    &               PROMPT          \\
18/10/12        &    56219.32 &      13.82   (0.05)  &       14.37   (0.05)          &       14.20   (0.05)  &       -                               &       -                       &               UVOT            \\
20/10/12        &       56220.89        &       13.84   (0.03)       &       14.45   (0.06)       &       14.28   (0.04)       &       14.09   (0.04)       &       13.99   (0.05)    &               RATCAM          \\
20/10/12        &    56220.96 &      13.88   (0.04)  &       -                               &       -                               &       -                               &       -                       &               UVOT            \\
20/10/12        &    56221.04 &      13.90   (0.05)  &       14.47   (0.05)          &       14.30   (0.05)  &       -                               &       -                       &               UVOT            \\
20/10/12        &       56221.18        &       13.80   (0.04)  &       14.51   (0.01)  &       14.34   (0.01)  &       14.15   (0.01)       &       14.02   (0.01)    &               EFOSC2          \\
21/10/12        &       56221.91        &       13.89   (0.02)  &       14.51   (0.02)  &       14.35   (0.01)  &       14.15   (0.01)       &         -                      &               RATCAM          \\
22/10/12        &    56222.78 &      13.93   (0.05)  &       14.61   (0.05)          &       14.45   (0.05)  &       -                               &       -                       &               UVOT            \\
22/10/12        &       56222.92        &       14.07   (0.10)  &       14.68   (0.02)  &       14.41   (0.02)  &       14.19   (0.02)       &       14.15   (0.02)    &               RATCAM          \\
23/10/12        &       56223.07        &       -                               &       -                               &       14.40   (0.01)  &       14.22   (0.01)       &       14.15   (0.04)   &               PROMPT          \\
23/10/12        &       56223.25        &       14.13   (0.03)       &       14.59   (0.02)  &       14.44   (0.01)       &       14.24   (0.01)       &       14.12   (0.02)    &               EFOSC2  \\      
24/10/12        &       56224.87        &       14.20   (0.02)       &       14.74   (0.01)       &       14.48   (0.02)  &       14.29   (0.01)       &       14.22   (0.05)    &               RATCAM\\
24/10/12        &    56224.99 &      14.26   (0.06)  &       14.66   (0.03)          &       14.49   (0.06)       &       -                               &       -                       &               UVOT\\
25/10/12        &       56225.02        &       -                 &       14.79   (0.02)  &       14.54   (0.05)       &       14.29   (0.03)       &       14.22   (0.02)    &               PROMPT\\
24/10/12        &       56225.12 &      14.27   (0.05)  &       -                               &       -                               &       -                               &       -                       &               UVOT\\
25/10/12        &       56225.88        &       14.33   (0.09)  &       14.86   (0.02)  &       14.65   (0.01)       &       14.41   (0.01)       &       14.34   (0.03)    &               RATCAM\\
26/10/12        &    56226.73 &      14.62   (0.06)  &       14.92   (0.05)          &       14.69   (0.06)          &       -                               &       -                       &               UVOT\\
28/10/12        &    56228.86 &      14.97   (0.06)  &       -                               &       -                               &       -                               &       -                       &               UVOT\\
29/10/12        &       56229.05        &       -                               &       15.42   (0.05)       &       15.02   (0.03)       &       14.74   (0.04)       &       14.61   (0.04)    &               PROMPT\\
28/10/12        &    56229.21 &      15.03   (0.06)       &       15.23   (0.06)       &       15.00   (0.07)       &       -                               &       -                       &               UVOT\\
30/10/12        &    56230.87 &      15.38   (0.06)  &       15.57   (0.05)          &       15.13   (0.06)  &       -                               &       -                       &               UVOT\\
31/10/12        &       56231.03        &       -                               &       15.49   (0.03)       &       -                               &       -                               &       -                       &               PROMPT\\
01/11/12        &    56233.35 &      15.25   (0.06)  &       15.39   (0.06)          &       14.97   (0.06)  &       -                               &       -                       &               UVOT\\
03/11/12   &    56234.98 &      15.17   (0.05)  &       15.22   (0.05)          &       14.78   (0.06)  &       -                               &       -                       &               UVOT\\
05/11/12        &       56236.09        &       14.87   (0.01)  &       15.21   (0.01)       &       14.92   (0.06)  &       14.71   (0.01)       &       14.36   (0.04)    &               EFOSC2\\
05/11/12   &    56236.71 &      15.02   (0.05)  &       15.17   (0.06)       &       14.81   (0.05)  &       -                               &       -                       &               UVOT             \\
06/11/12        &       56237.07        &       -                               &       15.32   (0.03)  &       -                               &       -                               &       -                       &               PROMPT\\
07/11/12        &       56238.03        &       -                               &       15.37   (0.04)       &       15.07   (0.03)  &       14.84   (0.03)       &       14.68   (0.03)    &               PROMPT\\
07/11/12   &    56238.55 &      15.39   (0.05)  &       15.40   (0.04)  &       15.15   (0.07)  &       -                               &       -                       &               UVOT\\
08/11/12        &       56239.02        &       -                               &       15.56   (0.02)  &       -                               &       -                               &       -                       &               PROMPT\\
08/11/12        &       56239.14        &       15.33   (0.01)       &       15.56   (0.01)  &       15.26   (0.01)       &       14.96   (0.02)       &       14.80   (0.02)    &               EFOSC2\\
09/11/12   &    56240.55 &      15.68   (0.05)  &       15.74   (0.05)  &       15.29   (0.06)       &       -                               &       -                       &               UVOT \\
11/11/12        &       56242.04        &       -                               &       15.99   (0.05)  &       15.50   (0.02)       &       15.21   (0.02)       &       15.03   (0.04)    &               PROMPT\\
11/11/12   &    56242.79 &      16.11   (0.07)  &       16.14   (0.06)  &       15.57   (0.06)       &       -                               &       -                       &               UVOT\\
11/11/12        &       56242.80        &       16.08   (0.05)  &       16.09   (0.02)  &       15.60   (0.05)       &       15.26   (0.07)       &       15.03   (0.05)    &               RATCAM\\ 
12/11/12        &       56243.04        &       -                               &       16.13   (0.04)       &       15.59   (0.02)       &       15.24   (0.03)       &       15.11   (0.04)    &               PROMPT\\
13/11/12        &       56244.04        &       -                               &       16.17   (0.02)  &       15.67   (0.02)       &       15.35   (0.02)       &       15.15   (0.04)    &               PROMPT\\
13/11/12   &    56244.92 &      16.39   (0.07)  &       16.24   (0.06)  &       15.71   (0.06)       &       -                               &       -                       &               UVOT\\
13/11/12        &       56244.82        &       16.281  (0.09)       &       16.30   (0.06)  &       15.69   (0.02)       &       15.35   (0.08)       &       15.22   (0.03)    &               RATCAM\\
14/11/12        &       56246.11        &       -                               &       -                               &       -                               &       15.37   (0.02)  &       -                       &               EFOSC2 \\
15/11/12        &       56246.17        &       -                               &       16.29   (0.03)  &       15.70   (0.07)       &       15.42   (0.03)       &       15.22   (0.05)    &               PROMPT \\
15/11/12   &    56247.41 &      16.52   (0.08)  &       16.41   (0.05)  &       15.81   (0.07)       &       -                               &       -                       &               UVOT\\
17/11/12   &    56248.54 &      16.58   (0.06)  &       16.43   (0.06)  &       15.85   (0.07)       &       -                               &       -                       &               UVOT\\
18/11/12        &       56249.08        &       -                               &       16.43   (0.03)  &       15.80   (0.03)       &       15.48   (0.02)       &       15.27   (0.04)    &               PROMPT\\
19/11/12   &    56250.54 &      16.87   (0.09)  &       16.59   (0.07)  &       15.96   (0.07)       &       -                               &       -                       &               UVOT\\
21/11/12        &       56252.13        &       -                               &       16.76   (0.03)  &       16.04   (0.06)       &       15.69   (0.04)       &       15.45   (0.06)    &               PROMPT\\
21/11/12        &       56252.16        &       17.12   (0.02)  &       16.77   (0.01)  &       16.21   (0.06)       &       15.73   (0.04)       &       15.48   (0.02)    &               EFOSC2\\
21/11/12   &    56253.08 &      17.15   (0.09)  &       16.82   (0.08)  &       16.15   (0.08)       &       -                               &       -                       &               UVOT\\
23/11/12  &     56254.58  &     17.51   (0.09)          &       16.87   (0.07)  &       16.29   (0.07)       &       -                               &       -                       &               UVOT\\
25/11/12        &       56256.14        &       -                               &       16.95   (0.05)  &       16.25   (0.04)       &       15.81   (0.04)       &       15.52   (0.05)   &               PROMPT\\
25/11/12  &     56257.17  &     17.64   (0.08)          &       17.04   (0.07)  &       16.38   (0.09)  &       -                               &       -                       &               UVOT\\
27/11/12  &     56259.03  &     17.66   (0.08)          &       17.19   (0.08)  &       16.47   (0.07)       &       -                               &       -                       &               UVOT\\
29/11/12        &       56260.83        &       17.64   (0.09)          &       17.19   (0.07)          &       16.52   (0.07)  &       -                               &       -                       &               UVOT\\
30/11/12        &       56261.15        &       -                               &       17.06   (0.04)  &       16.37   (0.03)  &       -                               &       -                       &               PROMPT\\        
01/12/12        &       56262.70        &       17.62   (0.08)          &       17.18   (0.07)          &       16.47   (0.07)       &       -                               &       -                       &               UVOT\\
03/12/12        &       56264.08        &       -                               &       17.22   (0.04)  &       16.40   (0.01)       &       15.93   (0.02)       &       15.62   (0.05)    &               PROMPT   \\
03/12/12        &       56264.82        &       17.46   (0.10)       &       17.14   (0.07)  &       16.48   (0.04)  &       15.86   (0.08)       &       15.70   (0.05)    &               RATCAM  \\
03/12/12        &       56265.06        &       17.66   (0.10)       &       17.31   (0.06)  &       16.54   (0.04)       &       16.06   (0.03)       &       15.66   (0.05)    &               EFOSC2\\
03/12/12  &     56265.11  &     17.82   (0.09)  &       17.37   (0.08)  &       16.69   (0.07)  &       -                               &       -                       &               UVOT\\
05/12/12        &       56266.65        &       18.01   (0.12)  &       17.68   (0.09)          &       16.72   (0.08)  &       -                               &       -                       &               UVOT\\
07/12/12        &       56269.16        &       18.59   (0.17)          &       18.16   (0.11)          &       17.04   (0.11)  &       -                               &       -                       &               UVOT\\
09/12/12        &       56270.85        &       18.87   (0.24)          &       18.31   (0.14)          &       17.32   (0.13)  &       -                               &       -                       &               UVOT\\
11/12/12        &       56272.08 &      -                               &       18.34   (0.04)  &       17.30   (0.04)  &       16.62   (0.03)       &       16.32   (0.09)    &               PROMPT\\
11/12/12        &       56272.76        &       18.97   (0.21)          &       18.52   (0.14)          &       17.46   (0.14)  &       -                               &       -                       &               UVOT\\
11/12/12        &       56273.05        &                                       &       -                               &       -                               &       16.72   (0.05)       &       -                       &               EFOSC2          \\
13/12/12        &       56274.62        &       19.06   (0.42)          &       18.86   (0.33)          &       18.06   (0.30)       &       -                               &       -                       &               UVOT\\
14/12/12        &       56275.07        &       -                               &       19.03   (0.06)  &       18.00   (0.05)  &       17.22   (0.04)       &       16.94   (0.07)    &               PROMPT\\
15/12/12        &       56276.73        &        -                              &       19.30   (0.22)          &       18.53   (0.21)  &       -                               &       -                       &               UVOT\\
17/12/12        &       56278.05        &       -                               &       19.16   (0.10)  &       18.35   (0.14)  &       17.46   (0.11)       &       17.24   (0.11)    &               PROMPT\\
20/12/12        &       56282.09        &       19.452  (0.10)  &       19.22   (0.09)  &       18.27   (0.06)  &       17.42   (0.05)       &       17.15   (0.10)    &               EFOSC2 \\       
29/12/12        &       56290.07        &       -                               &       $>$18.0                   &       18.55   (0.11)  &       -                               &       -                       &               PROMPT\\
\hline

\end{longtable}
\twocolumn

\begin{table*}
  \caption{UV photometry of SN 2009ip from {\it Swift}+UVOT, together with errors in parentheses. All values are in the UVOT system.}
  \label{tab:swift}
  \begin{tabular}{llccc}
  \hline
Date	   	 &	MJD		& 	uvw2		&	uvm2			&	uvw1  		\\
\hline
\hline
16/10/12   &	56216.71    &	13.39 	 (0.05) 	&     13.12 	 (0.05)	&    13.33	  (0.05)  \\
16/10/12   &	56216.87    &	13.54	 (0.05)	&     	 -		 	  	&    	- 	   	  \\
16/10/12   &   	56216.90    &   	-	 	 	 	&     13.29 	 (0.06)	&    	 -	   	    \\
18/10/12   &	56219.32    &	14.20	 (0.05)	&     13.97 	 (0.06)	&    13.79	  (0.06)  \\
20/10/12   &   	56220.93    &   	14.58	 (0.05)	&     	-  		 	  	&    	 -	   	  \\
20/10/12   &	56221.04    &	14.59	 (0.06)	&     14.39 	 (0.06)	&    14.10	  (0.06) \\ 
22/10/12   &   	56222.78    &   	14.90	 (0.06)	&     14.55 	 (0.06)	&    14.26	  (0.06)  \\
22/10/12   &	56222.98    &	-	 	 	 	&     14.56 	 (0.07)	&     -		       \\
22/10/12   &	56222.99    &	14.70 	 (0.05) 	&     -	  		 	  	&    	- 	   	  \\
24/10/12   &   	56224.99    &   	15.29 	 (0.07) 	&     14.93 	 (0.06)	&    14.48	  (0.06) \\ 
24/10/12   &	56225.00    &	-	  	 	  	&     15.15 	 (0.07)	&    	 -	  	     \\
24/10/12   &	56225.14    &	15.31 	 (0.06) 	&     	  -		 	  	&    	 -	  	   \\
26/10/12   &	56226.73    &	15.54 	 (0.07) 	&     15.26 	 (0.07)	&    14.85	  (0.06)  \\
28/10/12   &   	56229.21    &   	15.98 	 (0.08) 	&     15.61 	 (0.07)	&    15.39	  (0.07)  \\
30/10/12   &   	56230.87    &   	16.51 	 (0.09) 	&     16.00 	 (0.05)	&    15.59	  (0.07)  \\
01/11/12   &	56233.35    &	16.64 	 (0.08) 	&     16.19 	 (0.07)	&    15.76	  (0.07)  \\
02/11/12   &  	56234.28    &  	16.64 	 (0.08) 	&     16.27 	 (0.04)	&    -		  	       \\
03/11/12   &	56234.98    &	16.68 	 (0.08) 	&     16.20 	 (0.06)	&    15.64	  (0.06)  \\
05/11/12   &  	56236.71    &  	16.68 	 (0.07) 	&     16.42 	 (0.06)	&    15.65	  (0.06) \\
07/11/12   &  	56238.55    &  	16.51 	 (0.09) 	&     16.20 	 (0.07)	&    15.58	  (0.08) \\
09/11/12   &	56240.55    &	17.19 	 (0.06) 	&     16.74 	 (0.06)	&    16.06	  (0.08) \\
11/11/12   &  	56242.79    &  	17.57 	 (0.10) 	&     16.91 	 (0.09)	&    16.45	  (0.08) \\
13/11/12   &  	56244.92    &  	17.81 	 (0.10) 	&     17.51 	 (0.10)	&    17.08	  (0.16) \\
15/11/12   &  	56247.41    &  	18.00 	 (0.15) 	&     17.70 	 (0.13)	&    17.26	  (0.10) \\
17/11/12   &  	56248.54    &  	18.15 	 (0.12) 	&     17.80 	 (0.11)	&    17.28	  (0.10) \\
19/11/12   &  	56250.54    &  	18.43 	 (0.11) 	&     17.98 	 (0.12)	&    17.35	  (0.12) \\
21/11/12   &	56253.08    &	18.94 	 (0.17) 	&     18.68 	 (0.15)	&    17.57	  (0.11) \\
23/11/12   &	56254.58    &	19.19 	 (0.16) 	&     18.75 	 (0.10)	&    17.75	  (0.11) \\
25/11/12   &	56257.17    &	19.45 	 (0.22) 	&     19.06 	 (0.17)	&    18.07	  (0.10) \\
27/11/12   &	56259.03    &	19.68 	 (0.22) 	&     19.26 	 (0.18)	&    18.22	  (0.11) \\
29/11/12   &	56260.83    &	20.08 	 (0.26) 	&     19.14 	 (0.13)	&    18.36	  (0.11) \\
01/12/12   &	56262.70    &	20.10 	 (0.26) 	&     19.59 	 (0.17)	&    18.72	  (0.13) \\
03/12/12   &	56265.11    &	20.10 	 (0.26) 	&     19.56 	 (0.20)	&    18.68	  (0.14) \\
\hline

\end{tabular}
\end{table*}

\begin{table*}
\caption{NIR photometry of SN 2009ip in the 2MASS system. Errors are given in parentheses.}
\label{tab:nir}
\begin{tabular}{llcccr}
\hline
Date	   	 &	MJD		 	& 	J      			& 	H 			&	K			&	Instrument\\
\hline
\hline
12/10/12 &        56212.24  &      13.98        (0.07)        &  13.78 (0.04)           &     -               &      REMIR          \\   
19/10/12 &        56330.16	&	-				& 13.49 (0.07)		& -		& LUCIFER		\\
20/10/12 &	56220.83	&	13.99	(0.08)	& 13.67	(0.04)	&13.27	(0.08)	&	NOTCAM		\\
21/10/12 &        56221.02  &      14.19        (0.13)        &      -                           &                -                     &      REMIR           \\   
22/10/12 &	56223.19	&	14.17	(0.02)	& 13.71	(0.04)	&13.66	(0.13)	&	SOFI		\\   
26/10/12 &        56226.06  &      14.40        (0.12)        &        	-               &             -                     &      REMIR           \\  
05/11/12 &	56237.05	&	14.41	(0.08)	& 14.22	(0.11)	&13.94	(0.12)	&	SOFI		\\
11/11/12 &        56242.18 &        15.18        (0.08)        & 14.97   (0.10)         &     -                              &      REMIR           \\
13/11/12 &	56245.06	&	14.95	(0.07)	& 14.75	(0.08)	& 14.57	(0.06)	&	SOFI		\\
04/12/12 &	56266.02	&	15.33	(0.04)	& 15.09	(0.08)	& 14.91	(0.05)	&	SOFI		\\
21/12/12 &	56283.08	&	16.88	(0.11)	& 16.62	(0.08)	& 16.29	(0.09)	&	SOFI		\\
 
\hline 
 
\end{tabular}
\end{table*}

\end{document}